\documentclass[10pt]{article}
\usepackage{graphicx}
\usepackage{epsf}
\usepackage{epsfig}
\usepackage{amssymb}
\usepackage{amsmath}
\makeatletter
\@addtoreset{equation}{section}
\makeatother

\def \uqsl2 {$U_q$(sl$_2$) }
\def\bea{\begin{eqnarray}}
\def\eea{\end{eqnarray}}
\def\be{\begin{equation}}
\def\ee{\end{equation}}

\begin{document}

\def\wgta#1#2#3#4{\hbox{\rlap{\lower.35cm\hbox{$#1$}}
\hskip.2cm\rlap{\raise.25cm\hbox{$#2$}}
\rlap{\vrule width1.3cm height.4pt}
\hskip.55cm\rlap{\lower.6cm\hbox{\vrule width.4pt height1.2cm}}
\hskip.15cm
\rlap{\raise.25cm\hbox{$#3$}}\hskip.25cm\lower.35cm\hbox{$#4$}\hskip.6cm}}

\def\wgtb#1#2#3#4{\hbox{\rlap{\raise.25cm\hbox{$#2$}}
\hskip.2cm\rlap{\lower.35cm\hbox{$#1$}}
\rlap{\vrule width1.3cm height.4pt}
\hskip.55cm\rlap{\lower.6cm\hbox{\vrule width.4pt height1.2cm}}
\hskip.15cm
\rlap{\lower.35cm\hbox{$#4$}}\hskip.25cm\raise.25cm\hbox{$#3$}\hskip.6cm}}

\def\begeqar{\begin{eqnarray}}
\def\endeqar{\end{eqnarray}}

%
%
%
%
\noindent  SPhT 06/154
\begin{center}

\Large{Associative-algebraic approach to logarithmic conformal field
theories}
\vskip 1cm

{\large N. Read $^{a}$ and Hubert Saleur $^{b,c}$} \vspace{1.0em}

{\sl\small $^a$
Department of Physics\\
Yale University, P.O. Box 208120, New Haven, CT 06520-8120 USA\\}
{\sl\small $^b$ Service de Physique Th\'eorique, CEA Saclay,\\
Gif Sur Yvette, 91191, France\\}
{\sl\small $^c$ Department of Physics and Astronomy,
University of Southern California\\
Los Angeles, CA 90089, USA\\}

\end{center}

\begin{abstract}
We set up a strategy for studying large families of logarithmic
conformal field theories by using the enlarged symmetries and
non--semi-simple associative algebras appearing in their lattice
regularizations (as discussed in a companion paper). Here we work
out in detail two examples of theories derived as the continuum
limit of XXZ spin-$1/2$ chains, which are related to spin chains
with supersymmetry algebras gl($n|n$) and gl($n+1|n$),
respectively, with open (or free) boundary conditions in all
cases. These theories can also be viewed as vertex models, or as
loop models. Their continuum limits are boundary conformal field
theories (CFTs) with central charge $c=-2$ and $c=0$ respectively,
and in the loop interpretation they describe dense polymers and
the boundaries of critical percolation clusters, respectively. We
also discuss the case of dilute (critical) polymers as another
boundary CFT with $c=0$. Within the supersymmetric formulations,
these boundary CFTs describe the fixed points of certain nonlinear
sigma models that have a supercoset space as the target manifold,
and of Landau-Ginzburg field theories. The submodule structures of
indecomposable representations of the Virasoro algebra appearing
in the boundary CFT, representing local fields, are derived from
the lattice. A central result is the derivation of the fusion
rules for these fields.

\end{abstract}

\begin{center}{}\end{center}

\bigskip

\section{Introduction}

The interest in logarithmic conformal field theories (CFTs) (see
\cite{Lectures} for an introduction) has grown over the last ten
years or so as their potential role in condensed matter as well as
string theory applications has become more evident. Condensed
matter applications include critical geometrical models in two
dimensions such as percolation and polymers (see \cite{Flohr} for
a recent contribution and review), and critical points in
non-interacting disordered fermion models in $2+1$ dimensions (see
\cite{glr,Leclair,Caux} for recent contributions), such as the
transition between plateaux in the integer quantum Hall and spin
quantum Hall effect. String theory applications include the study
of sigma models on non-compact (super-) target spaces such as
PSU$(1,1|2)$ \cite{Volker}. Logarithmic CFTs are also fascinating
mathematical objects in their own right, bound to attract the
attention of representation theory experts.

Logarithmic CFTs were encountered as early as 1987 by Knizhnik
\cite{Knizhnik}. In 1992, Rozansky and Saleur tackled in some
details one of the simplest such theory, the WZW model on the
U($1|1$) supergroup. They found that the non-semisimplicity of
U($1|1$) carried over to the CFT, with the possibility of a
non-diagonal Virasoro generator (here, proportional to the
superalgebra Casimir) $L_{0}$, indecomposable operator product
expansions (OPEs), and logarithms in some of the correlators
\cite{RozSal}. The potential generality of these features was
realized almost simultaneously by Gurarie in 1993 \cite{Gurarie1},
who coined the name logarithmic CFT.

Certainly the theory of symplectic fermions is the best understood
logarithmic CFT \cite{Kausch}. [A close second is the WZW model on
U($1|1$) where powerful results were obtained recently  using the
minisuperspace approach \cite{SchomSal}. It is not clear to what
extent the current algebra structure makes this example qualitatively
different from the logarithmic CFTs of more physical interest.] A
natural extension of the symplectic fermions has given rise to the
so-called ``augmented $c_{p,1}$ models'', which exhibit an enlarged
triplet ${\cal W}$ algebra \cite{Kausch,Flohri,Feigini}. An essential
feature of these theories is that the identity has a logarithmic
partner, and thus that there are at least two fields of vanishing
dimension. There have been some attempts to generalize the structure
of the augmented $c_{p,1}$ models to other series, in particular the
$c_{p,2}$ case \cite{Fjelstad,Flohrii}.

One of the most interesting cases, with numerous condensed matter
applications to disordered systems, corresponds to vanishing
central charge. Consideration of singularities in OPEs as
$c\rightarrow 0$ led Gurarie to the conjecture that in these
theories the stress energy tensor $T$ had a logarithmic partner
$t$ such that $L_{0}$ had a Jordan cell of dimension 2 at the
level $h=2$ \cite{Gurarie2}. In this and in subsequent papers
Gurarie and Ludwig \cite{GurarieLudwig1,GurarieLudwig2} proposed
to use the modes of $t$ to define an extension of the Virasoro
algebra characterized by a parameter $b$. [Although appealing
intuitively and successful in some cases, the idea has not borne
much fruit, in particular because of considerable difficulties in
putting together chiral and non chiral sectors.] The problem has
also been discussed from a slightly different point of view of the
replica limit by Cardy \cite{Cardy}.

Most of the approaches mentioned so far start from the
consideration of abstract conformal field theories. Such
generality may not be the easiest way to proceed. In 2001, we
defined families of supersymmetric lattice models, whose continuum
limits we identified with the IR limits of conformal
supersymmetric non-linear sigma models, or in some cases with
Landau-Ginzburg theories \cite{ReadSal}. [For instance, one of
these lattice models was defined using alternating fundamental and
conjugate representations of sl($n+1|n$) and argued, in the
continuum limit, to be described by the IR limit of the $CP^{n|n}$
model at $\theta=\pi$. This model is closely related to the
properties of hulls of percolation \cite{glr}. A similar model was
defined for dilute and dense polymers.] In all these cases, we
obtained the full spectrum of the associated logarithmic conformal
field theory, and found considerable richness, including an
infinite set of Virasoro primary fields which all have rational
conformal weights, and indications of a large non chiral symmetry
algebra commuting with Virasoro. We note that in all our examples
with $c=0$, there was a single field with vanishing dimension. We
expect in general that the logarithmic theories we uncovered do
not coincide with any of the extended $c_{p,1}$ or $c_{p,2}$
models mentioned before (except for the symplectic fermions).

It is worthwhile to consider a broad view of the strategies
pursued in these various references. First, we point out that the
traditional idea of a {\em rational} CFT contains two essential
ingredients \cite{ms}: the local fields (or corresponding states)
fall into a {\em finite} number of representations of the chiral
algebra (which in the strictest view of rational CFT is generated
by a finite number of bosonic local fields of integer conformal
weights), and these representations are semisimple, that is fully
decomposable as direct sums of irreducible representations of the
chiral algebra. Hence also the operator products of local fields
decompose fully into sums of these local fields. Next, we note
that the occurrence of logarithms in correlators is due to $L_0$
not being diagonalizable, and this results from the appearance of
fields transforming in non-semisimple representations of the
Virasoro algebra. One is then led to consider indecomposable
representations that are not necessarily irreducible. The next
problem is how to control these theories that do not satisfy this
requirement of traditional rational CFTs. Many of the papers cited
above try to do so by considering extended chiral algebras
\cite{ms}, such that the number of indecomposable representations
is finite, so that in this sense one may speak of ``rational
logarithmic CFTs''. But many applications, such as to disordered
systems, and all the theories we consider in this paper, appear to
possess only the Virasoro algebra as the chiral algebra, and
infinitely many (non-isomorphic) indecomposable representations
appear as local fields. In addition, these theories may even fail
to be quasirational, that is infinite sums of local fields may be
produced in ope's. In these irrational CFTs, it is not clear how
the analysis is to be organized.

We propose that the use of symmetry algebras that commute with the
{\em full} chiral algebra \cite{ReadSal2} can be a powerful
organizing principle, even when semisimplicity of the algebraic
structures is lost. These symmetries and structures can be
usefully studied by starting with lattice models. The idea of
gaining understanding of logarithmic CFT (LCFT) by studying
lattice realizations in details has been around for a while, and
put forward most recently (albeit without the algebraic
connotations we consider crucial) in \cite{Zuber}. It does open a
fruitful new route to progress. Indeed, the lattice equivalent of
the reducible but indecomposable Virasoro representations in LCFT
is the non semi-simplicity of the various algebras underlying the
lattice models. [Examples include the Temperley-Lieb (TL) algebra
or $U_q($sl$_2)$, neither of which is semi-simple when $q$ is a
root of unity; such values are usually those of greatest
interest.] A great deal of progress has been made in the
mathematics literature in this area over the last fifteen years or
so. Our strategy will be to exploit this progress and infer from
it results about the LCFTs.

More precisely, the lattice models we deal with all have a similar
structure (we discuss here only the boundary case, even though
many results generalize to the case of the bulk). The local
Hamiltonian densities generate an associative algebra (in our
examples, the TL algebra) whose commutant, in the lattice Hilbert
space of interest, is $U_q$(sl$_2$) in the example of the spin-1/2
chain, or in the supersymmetric models in Refs.\
\cite{ReadSal,ReadSal2} a remarkably large algebra whose
representations are in ``Morita equivalence'' with those of
$U_q$(sl$_2$). Using results from the mathematics literature
\cite{martinbook,Martini,Martinii,Jonesi} and others we derive
\cite{ReadSal2}, we can analyze in full detail the structure of
these two commuting algebras (which are in so-called Ringel
duality) and their representations in the lattice model. It is
then a natural step to conjecture the corresponding structure in
the continuum limit: the algebra of Hamiltonian densities goes
over to the Virasoro algebra \cite{KooSal,FevePear}, while the
commutant goes over to an algebra of non-local charges. The
general structure of the theory is one in which the Virasoro
algebra commutes with a certain symmetry algebra, but the latter
is {\em not} any kind of current algebra, and nor is it a Yangian.
The structure of indecomposable representations in the lattice
model goes over to the continuum limit, and we thus obtain
information about the representations of the Virasoro algebra in
our logarithmic theories. These representations are typically
reducible but indecomposable with a diamond (or quartet) shape and
made out of two ``standard modules'' connected by some ``glue''.
They bear a lot of resemblance to the so-called staggered modules
introduced abstractly in \cite{Rho}. The full structure of the
Hilbert space involves big indecomposables representations of
Virasoro and of its commutant, and can be exhibited in the form of
what we will call a ``staircase diagram''.

Although our approach does not so far yield much information on
the detailed structure of the Virasoro indecomposable
representations, it is enough to give access to the fusion rules
for our logarithmic theories. We work these out in detail in some
cases.

The remainder of this paper is organized as follows. Section 2
describes the general framework of the lattice models, algebraic
structures, and the continuum limit, for $q$ not a root of unity,
and also the general features of cases in which $q$ is a root of
unity. The following three sections go into details about
particular examples of special interest. Each one begins with a
summary of the physical systems considered, viewed as loop models,
or as field theories of certain nonlinear sigma models or
Landau-Ginzburg theories, then describes the algebraic structures
leading to the continuum limit, and ending with the fusion rules.
Section 3 is the case $q=i$ (or central charge $c=-2$ in the
continuum limit), which corresponds to symplectic fermions or
dense polymers. Section 4 is the case $q=e^{i\pi/3}$ ($c=0$ in the
continuum limit), which corresponds to percolation, the spin
quantum Hall transition, or a certain nonlinear sigma model at
$\theta=\pi$. Section 5 is the case of dilute polymers ($c=0$), or
of the critical point in a certain nonlinear sigma model or
Landau-Ginzburg theory. For this theory we discuss the continuum
directly, and not the lattice models. Finally, section 6 is a more
physical discussion of the structures for the cases with $c=0$,
using operator product expansions, and making contact with the
work of Gurarie and Ludwig \cite{GurarieLudwig1,GurarieLudwig2}.

To make reading easier,  we provide a list of notations for some
of the algebraic objects we introduce, all of which implicitly
depend on $q$ or the central charge $c$, and also on length $L$
for the TL algebra:

${\rm T}_{j}$: tilting modules over $U_q$(sl$_2$)

${\rm P}_{j}$: projective modules over TL$_{2L}(q)$

$r_{j}$: standard modules over the Virasoro algebra

$R_{j}$: simple (or irreducible) modules over the Virasoro algebra

${\cal R}_{j}$: projective modules over the Virasoro algebra


\section{Algebraic structure on the lattice and in the continuum}

What we will do here is (i) take the continuum limit of the lattice model
on a strip to obtain the representation content of the states, and (ii)
also use the lattice to infer fusion rules for the boundary CFT.

In the companion paper \cite{ReadSal2}, we showed that the
commutant algebras of the Temperley-Lieb algebra in some
supersymmetric lattice models are Morita equivalent
\cite{AndersonFuller} to $U_q$(sl$_2$). Accordingly, here we can
use the $U_q$(sl$_2$)-invariant spin-1/2 chain, secure in the
knowledge that the symmetry structure is the same in the
supersymmetric constructions that apply when $m=q+q^{-1}$ is an
integer. We often refer to representations as modules, as these
are essentially the same things for our purposes (however, in the
more mathematical discussions, saying ``representation'' in place
of ``module'' would be quite misleading in some places, especially
when preceded by ``projective''). Then an irreducible
representation is the same thing as a simple module, and it is
also common to speak of modules being ``over'' the algebra that
they represent. We assume familiarity with basic algebraic
concepts such as sub- and quotient- objects, here usually for
modules. For a general reference on the algebraic background, we
suggest Ref.\ \cite{AndersonFuller}.


\subsection{XXZ spin-1/2 chain, TL algebra and continuum limit: generic case}

The Hamiltonian of the $U_q$(sl$_2$)-invariant spin-1/2 chain is
essentially the XXZ chain, but with some boundary terms (here the
$\sigma$ operators are Pauli matrices):%
\begin{equation}
H=\frac{1}{2}\sum_{i=0}^{n-2}
\left(\sigma_{i}^{x}\sigma_{i+1}^{x}+\sigma_{i}^{y}\sigma_{i+1}^{y}+
\frac{q+q^{-1}}{2}
\sigma_{i}^{z}\sigma_{i+1}^{z}\right)+\frac{q-q^{-1}}{2}\left(\sigma_{0}^{z}
-\sigma_{n-1}^{z}\right)
\end{equation}
The Hamiltonian $H$ is built from generators $e_i$,
\begin{equation}
e_i = \frac{q+q^{-1}}{4}-\frac{1}{ 2}\left(\sigma_{i}^{x}
\sigma_{i+1}^{x}+\sigma_{i}^{y}\sigma_{i+1}^{y}\right)-
\frac{q+q^{-1}}{4}
\sigma_{i}^{z}\sigma_{i+1}^{z}-\frac{q-q^{-1}}{4}\left(\sigma_{i}^{z}-\sigma_{i+1
}^{z}\right)
\end{equation} %
of the TL algebra TL$_n(q)$, which is defined as the algebra
generated by a set of elements $e_i$, $i=0$, $1$, \ldots, $n-2$,
subject to the relations
\begin{eqnarray}
e_i^2&=&me_i,\\
e_ie_{i\pm1}e_i&=&e_i,\\
e_ie_{i'}&=&e_{i'}e_i\label{tlrels}\end{eqnarray}%
and no other relations algebraically independent of these.  These
are satisfied by the above expressions in the spin-1/2 chain. Then
$H=-\sum_{i=0}^{n-2}e_i$ (up to an irrelevant additional
constant). All future references to the spin-1/2 or XXZ chain are
to this model.

First we briefly summarize the case of $q$ generic (i.e., $q$ not
a root of unity) and its continuum limit. For $q$ not a root of
unity, the TL algebra is semisimple and has irreducible
representations (all such will be referred to as simple modules
hereafter) labeled by $j=0$, $1/2$, $1$, \ldots, $n/2$ but
restricted to values such that  $n/2+j$ is an integer.
The dimensions of these simple modules are $d_j$,%
\begin{equation}
d_j=\left(\begin{array}{c}
n\\
n/2+j\end{array}\right)-\left(\begin{array}{c}
n\\
n/2+j+1\end{array}\right).
\end{equation}
These modules are referred to as standard (or Specht) modules.

The commutant of the TL algebra in the spin-1/2 chain is a
finite-dimensional homomorphic image of $U_q$(sl$_2$), that we
will denote $U_q($sl$_2)^{(n)}$ (it is also called the $q$-Schur
algebra). For $q$ generic, its simple modules are again labeled
$j=0$, $1/2$, $1$, \ldots, $L/2$, where $n/2+j$ is an integer.
Throughout the paper, we will use the familiar term ``spin'' to
refer to $j$, unless otherwise noted. These modules have
dimensions $D_j=2j+1$, independent of $n$ except that $j\leq n/2$.
They are called Weyl modules [or standard modules of
$U_q$(sl$_2$)] (we can drop the superscript $n$ as they are
independent of it, and the limit $n\to\infty$ can be taken for
these). They are the deformations to $q\neq 1$ of the usual
irreducibles of SU(2).

If two chains of lengths $n_1$, $n_2$ are joined end to end (with $n_1$ on
the left), one obtains fusion rules for both algebras. The states of a
single chain of length $n_1$ fall into multiplets (simple modules) of the
tensor product of commuting algebras, TL$_{n_1}(q)\otimes
U_q($sl$_2)^{(n_1)}$ which are labeled by $j$; these modules are the
tensor product of the corresponding simple modules of the two
respective algebras. When the two chains are joined, the resulting chain
of length $n=n_1+n_2$ can be analyzed in the same way. If the standard
modules of $U_q$(sl$_2$) are denoted $V_j$, then tensor products of such
modules (where the tensor product is defined as in the usual way for
vector spaces) decompose fully into a direct sum,%
\begin{equation}
V_{j_1}\otimes V_{j_2}=\bigoplus_{j=|j_1-j_2|}^{j_1+j_2}V_j,\end{equation}%
exactly as for SU(2). The standard modules of TL must behave in a
similar way, but now the product operation is defined using
induction (which may be familiar from induced representations in
group representation theory) from the product space as a module
over TL$_{n_1}(q)\otimes $TL$_{n_2}(q)$ to a module over
TL$_{n}(q)$, which contains the former as a subalgebra. The
general theory described in Ref.\ \cite{ReadSal2} implies that
indeed the ``fusion rule'' for this definition of a product of TL
modules must take the same form as that above for $U_q$(sl$_2$)
(note this product is a functor, and not just a decomposition of
the tensor product of vectors; in particular the dimensions of the
tensor product of the modules on the left and the sum of those on
the right are not equal).

Now we pass to the continuum limit. From this point on, we
consider only $n=2L$ even for the time being. (This does no harm
as the modules with $j$ integer close on themselves under
products, as we see in the fusion rules above, and will see for
non-generic $q$ below. This restriction corresponds to the
``oriented loops'' models in ref.\ \cite{ReadSal2}.) In the
continuum limit we consider only the Fourier modes of the $e_i$s
of wavevectors $k$ with $kL$ fixed, and also consider only states
spanned by eigenstates with energies $E$ (measured from the ground
state energy) with $EL$ fixed. Then the Fourier modes of the
$e_i$s become the Virasoro generators $L_n$, $n=\ldots$, $-2$,
$-1$, $0$, $+1$, \ldots, \cite{KooSal}. The standard modules of
the TL algebra, labeled by $j$, become certain modules of the
Virasoro algebra, which we will also call standard, and which are
still labeled by $j$. These modules are highest weight modules,
that is they are generated by the action of the $L_n$s on the
state of lowest energy or of lowest $L_0$ eigenvalue in the
module. They are not, however, Verma modules, which are
constructed from a highest weight state by lowering by $L_n$s with
$n<0$, and using only the commutation relations of the Virasoro
algebra. The conformal weight $h$ of the module, which is the
$L_0$ eigenvalue of the highest weight state, takes one of the
values $h_{r,s}$ at which a null vector (the highest weight in a
submodule) appears in the Verma module, which must be set to zero
in order to obtain a simple module. For generic values of the
central charge, there is only a single null vector, and only when
the conformal weight takes one of the values in the Kac table
\cite{cft,BPZ}%
\begin{equation}
h_{r,s}= \frac{[(x+1)r-xs]^{2}-1}{4x(x+1)},\label{kac}\end{equation}%
in which $x$ is determined by the central charge %
\begin{equation}
c =1-\frac{6}{ x(x+1)}.\label{c}\end{equation} %
The generic values of $c$ are
those where $x$ is irrational. When we take the limit of spin-1/2
chain, the resulting central charge follows from $q=e^{i\pi/
(x+1)}$. The conformal weights that occur are $h_{1,1+2j}$ in the
Kac table. In the corresponding Verma module, the null vector is
at conformal weight $h_{1,-1-2j}$, Hence the Virasoro character of
the simple modules that appear as direct summands in the
$L\to\infty$
limit of the chain are (the trace is over states in the simple module only)%
\begin{equation}
{\rm Tr}\,
\widehat{q}^{L_0}=\frac{\widehat{q}^{~h_{1,1+2j}}-\widehat{q}^{~h_{1,-1-2j}}}
{P(\widehat{q})},\end{equation} where $\widehat{q}$,
$|\widehat{q}|<1$ is a parameter, $\widehat{q}=e^{-\pi\beta/2 L}$,
where $\beta$ is the inverse temperature (the length in the
imaginary time direction), and $P(\widehat{q})$ is the inverse of
the Euler partition function, and is related to the Dedekind
$\eta$ function,
\begin{equation}
P(\widehat{q})=\prod_{r=1}^\infty(1-\widehat{q}^{~r})=
\widehat{q}^{~-1/24}\eta(\widehat{q}).
\end{equation}
Thus the partition function of the continuum limit of the XXZ
chain in the generic case is%
\be%
\lim {\rm Tr}\, e^{-\beta (H-E_0(L))}={\rm Tr}\,
\widehat{q}^{~L_0}= \sum_{j=0}^\infty
(2j+1)\frac{\widehat{q}^{~h_{1,1+2j}}-\widehat{q}^{~h_{1,-1-2j}}}
{P(\widehat{q})}.\ee %
Here the continuum limit is taken with $\widehat{q}$ fixed, after
subtracting the ground state energy $E_0(L)$ of the finite-$L$
chain from the Hamiltonian, and the trace is over all the states
of the spin-1/2 chain. [We emphasize that this is the ordinary
trace on the vector space, not a $q$-trace; it corresponds to what
were called the modified partition functions in Ref.\
\cite{ReadSal}.] If instead the subtraction were the length times
the ground state energy density $e_0(L)=\lim_{L\to\infty}
E_0(L)/L$ of the chain as $L\to\infty$, this expression would be
multiplied by $\widehat{q}^{~-c/24}$. This is the way we will
define the
partition function here and in the following:%
\be%
Z=\lim {\rm Tr}\, e^{-\beta (H-e_0(L)L)}={\rm Tr}\,
\widehat{q}^{~L_0-c/24}= \sum_{j=0}^\infty
(2j+1)\frac{\widehat{q}^{~h_{1,1+2j}-c/24}-\widehat{q}^{~h_{1,-1-2j}-c/24}}
{P(\widehat{q})}.\ee %
Let us note that this partition function can also be obtained by
field-theoretic (rather than algebraic) means, by bosonizing the
spin-1/2 chain or the 6-vertex model, that is, by representing it
by a scalar field with a background charge, and taking the
continuum limit \cite{sb,cardy}.

It is known that \cite{BPZ}, as a consequence of the null vectors
that were set to zero, the operator products of the fields
$\phi_j(z)$ that correspond to the states in the modules decompose
according to the fusion rules
\begin{equation}
\phi_{j_1}\times
\phi_{j_2}=\sum_{j=|j_1-j_2|}^{j_1+j_2}\phi_j.\end{equation} These
clearly correspond to those we inferred from the induction product
in the chains, or from the $U_q$(sl$_2$) symmetry of the whole
construction. In the following we use similar reasoning to infer
fusion rules also in the non-generic cases in which
representations are not fully decomposable.

\begin{figure}
\begin{center}
 \leavevmode
 \epsfysize=80mm{\epsffile{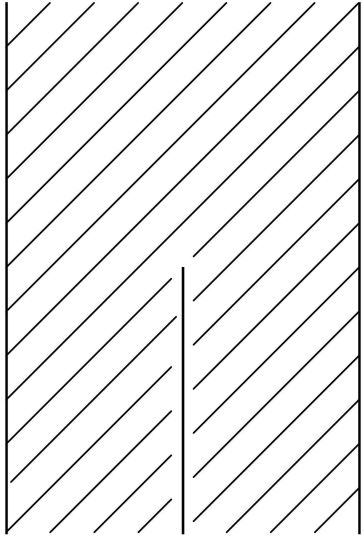}}
 \end{center}
 \protect\caption{The ``slit-strip'' picture of open string
 interactions, or of joining two chains end to end, with time
 increasing upwards. The shading indicates the interior
 of the domain.} \label{slitstrip}
\end{figure}

In this procedure, we have constructed the fusion product
algebraically by joining two chains end to end and taking the
continuum limit. One may wonder whether this really yields the
operator product of corresponding local boundary fields in the
CFT, which is usually pictured as bring points on the boundary
together. We will not discuss this fully, but provide some
heuristic arguments. If we view joining the chains as an event in
time, then it corresponds to a picture like that shown in Fig.\
\ref{slitstrip}, in which the continuum limit has been assumed.
(The chains are shown as equal in length, though more generally
the ratio of their lengths can be arbitrary, but is held fixed
when the continuum limit is taken.) If we wish to consider the
fusion of states that are $L_0$ eigenstates, for example, then the
two initial states may be prepared at early imaginary time $t\to
-\infty$, and the matrix element with a desired final state can be
taken, by preparing that state at $t\to+\infty$. As conformal
transformations move points within the interior of the domain (and
even deform the domain), it is natural that the tensor product of
two copies of Virasoro merges into a single copy of Virasoro, as a
proper subalgebra. This occurs because the continuum analog of the
extra generator $e_{n_1-1}$ must be included along with the
others. Then the corresponding actions of these algebras on the
states are related by the induction functor. Thus the use of the
induction functor for Virasoro is natural in the slit-strip
picture (i.e., in the operator formalism). The conventional
picture of the incoming and outgoing states as corresponding to
local fields at a point on the boundary, which are brought
together in an ope, is recovered by using a conformal mapping from
the slit-strip to the unit disk, or the upper half plane, with the
two incoming and one outgoing state(s) at time $\to\pm\infty$
represented by points on the boundary. This conformal mapping is
described for example in Ref.\ \cite{gsw}. Incidentally, it has
often been remarked in the CFT literature that the fusion product
is not simply a tensor product of representations of Virasoro (or
more generally, the chiral algebra). The fusion product defined as
the induction functor for joining strips end to end should agree
with the ``deformed tensor product'' defined in some of these
references \cite{ms}, at least for the present case of boundary
CFT.

The states or fields we find in this construction based on the TL
algebra at generic $q$ correspond to the first column of the Kac
table. The first row of the Kac table has identical fusion rules,
and so one would expect a lattice construction to lead to this
theory also. Such a model is the O($m$) loop model
\cite{nien,dfsz,nienrev} and its supersymmetric versions
\cite{ReadSal,ReadSal2} in which the critical theory (which exists
for $-2< m \leq 2$) is referred to as the dilute theory, and where
another algebra (essentially a two colour version of the TL
algebra) replaces the TL algebra. We emphasize that it has the
same symmetry  as the TL (dense loops) model, and that it too
comes in two versions, oriented loops (only integer $j\geq0$) and
unoriented loops (including all integer values of $2j\geq 0$).


\subsection{Non-generic cases}

When $q$ is a root of unity (but $q\neq\pm1$), we will define
$r>1$ to be the smallest integer such that $q^r=\pm 1$. The
structure of both TL and $U_q$(sl$_2$) changes, and this can be
traced to the vanishing of the $q$-integers
$[N]_q=(q^N-q^{-N})/(q-q^{-1})$ for some $N$ for such $q$ values:
$[N]_q=0$ for $N\equiv0$ (mod $r$)
\cite{PasquierSaleur,Martini,Martinii}.

For $q$ a root of unity, standard modules of both TL and \uqsl2
can be defined, and have the same dimensions $d_j$, $2j+1$,
respectively, as before. However, they are not all irreducible. It
turns out that for $j$ such that $[2j+1]_q=0$, they remain
irreducible, while for other $j$, they contain a proper submodule
but are indecomposable. For both algebras, the reducible standard
modules have the same structure: the largest proper submodule (or
radical) of the module is irreducible, and the quotient by this
submodule is also. Then we will represent the structure of such a
module by a diagram like
\be%
\begin{array}{ccc}
\circ&&\\
&\hskip-.2cm\searrow&\\
&&\hskip-.3cm\bullet\label{stanmod}
\end{array}. \ee%
Here the open circle represents the states in the simple quotient
module or ``top'' (or ``head''), and the closed circle the simple
submodule or ``foot''. The arrow represents the action of the
algebra; there is some element of the algebra that maps the top to
the foot, but not vice versa, as well as elements that map the top
into itself and the foot into itself. The positions of the circles
on the page is not particularly important, as the key information
is the arrows and (for these standard modules) whether a circle is
open or closed. Below, we will also represent the structure of
other modules by similar, though more complex, diagrams. In these,
each circle is a nonzero simple subquotient module, and arrows
show the action of the algebra other than within the simple
subquotients, with the convention that composites of arrows should
also be understood as present implicitly. In our examples, a head
and a foot can usually be defined uniquely, but there will be
other subquotients in between.

Later, the positions of the circles will be replaced by additional
information such as the spin or dimension of the simple module in
question. The general rule for the spin values can be given here:
For either \uqsl2 or TL, we label standard modules by $j=0$, $1$,
\ldots, $L$. We choose the labels for the simple modules to be
$j=0$, $1$, \ldots, $L$ also (or in some cases, a subset of
these). For reducible standard modules of $U_q($sl$_2)$, the
simple module at the {\em top} has the same value of $j$; the foot
is a simple module isomorphic to a simple module with {\em
smaller} $j$. For the TL algebra, it is the same, except that the
foot has a larger $j$ than the head.

When the XXZ chain is analyzed for $q$ a root of unity, it
decomposes under the TL algebra into a direct sum of TL modules.
When analyzed under $U_q($sl$_2)$, it likewise is a direct sum.
However, in neither case are the direct summands the standard
modules just described, except for those standard modules that
remain simple at this $q$, some copies of which appear as in the
generic cases. For the reducible standard modules, each algebra
action mixes them, mapping states from one standard to another
\cite{PasquierSaleur,martinbook,Martini,Martinii}. The direct
summand modules will be referred to as ``tilting'' modules, in the
case of either algebra. (We should point out that non-semisimple
algebras typically have infinitely many distinct, i.e.\ mutually
non-isomorphic, indecomposable modules, even when the algebra is
finite dimensional. It becomes important to explain {\em which}
particular types of indecomposable modules are of interest.) The
tilting modules turn out to have a fairly simple structure also,
consisting of two standard
modules with arrows connecting them, typically in the form:%
\be%
  \begin{array}{ccccc}
       &&\circ&&\\
       &\swarrow\hskip-.2cm&&\hskip-.2cm\searrow&\\
     \bullet\hskip-.3cm&&&&\hskip-.3cm \circ\\
       &\searrow\hskip-.2cm&&\hskip-.2cm\swarrow&\\
       &&\bullet&&
       \end{array}.\label{tiltmod}
       \ee%
Thus, the module is seen to contain a submodule that is a standard
module, consisting of an open and a closed circle together with an
explicit arrow that connects them. The quotient by this submodule
is also standard. The detailed structure of all these modules will
be exhibited for several important values of $q$ (or $m$) in the
following sections, as will the inter-relation of the commuting TL
and \uqsl2 actions.

In addition to the tilting modules, other important modules are
the projective modules. Projective modules may be defined as
occurring as a direct summand in some free module, or in several
other equivalent ways. We will be especially interested in the
indecomposable projective modules of the TL algebra. For finite
dimensional algebras, there is one indecomposable projective
module (up to isomorphism) corresponding to each distinct simple
module; the latter appears as the top (the unique simple quotient
module) of this projective module. This fact makes it fairly
straightforward to read off the indecomposable projective modules
from the structure of the XXZ chain as a direct sum, since the
algebra acts faithfully. In many cases, the tilting modules are
also indecomposable projective modules.

Joining chains of lengths $L_1$, $L_2$ end to end leads to natural
operations making a tensor product of modules over \uqsl2 into
another such module. These agree with the product operations
usually defined for \uqsl2 as a quantum group. This product
decomposes as a direct sum of indecomposable modules, and as
tilting modules are defined as direct summands in the XXZ chain,
the tensor product of tilting modules is again tilting. The fusion
rules for the tilting modules are then those of most basic
interest (and are determined by the decomposition of a product of
indecomposable tilting modules into indecomposables). Also, as
this product is just the ordinary tensor product operation of
vector spaces, the sum of the dimensions of the indecomposables
produced is the product of the dimensions of the modules we began
with. For $q$ a root of unity, the fusion rules are however
modified from those in the generic cases, and will be discussed in
the examples below.

In our previous paper \cite{ReadSal2}, we also explained how a
product is defined on the TL modules when joined end to end, based
on induction as mentioned earlier. In this case, the dimensions of
modules are not conserved, but there are again fusion rules.
Crucially, the module induced from a tensor product of projective
modules of the two short chains is a projective module of the
chain of length $L_1+L_2$, and can be decomposed as a direct sum
of indecomposable projective modules. Hence we can define fusion
rules for products of projective modules, which are again
determined by the decomposition of a product of indecomposable
projective modules into indecomposable projective modules. Most
importantly, the fusion rules for indecomposable projective
modules of TL are the {\em same} as those for indecomposable
tilting modules of $U_q($sl$_2)$ \cite{ReadSal2}.

We also discussed previously \cite{ReadSal2} how the limit
$L\to\infty$ can be taken, both at a purely algebraic level, and
focussing on the low-energy states as a continuum limit. The first
can be well-defined mathematically, while the second is presently
less rigorous and more heuristic. For the symmetry structure,
given by \uqsl2 or a Morita equivalent of it, the purely algebraic
construction is sufficient, as it yields a limit algebra, which is
\uqsl2 for the XXZ case. The dimensions of modules are stable as
$L$ increases, but additional modules appear. The fusion rules are
also given by the finite-dimensional analysis and remain the same
as $L$ increases [and in discussing \uqsl2 above, it was natural
to ignore the distinction between it and its finite-dimensional
quotients]. On the other hand, for the TL algebra, the dimensions
of corresponding modules increase with $L$, and there seems to be
no purely algebraic way to define the limiting algebra that is
suitable for our purposes \cite{ReadSal2}. Nonetheless, the fusion
rules behave stably (as they are controlled by the symmetry). In
the purely algebraic limit, the limit of a projective module is
still a projective object in the limiting category of modules. In
the continuum limit, all evidence suggests that the modules of TL
become Virasoro modules. The limits of projective modules will
still be referred to as projective, and continue to play an
important role. The standard (resp.,\ tilting) modules of TL
become standard (resp.,\ tilting) modules of Virasoro, also. We
note that the Virasoro algebra may not act faithfully on the
limiting space of states. This is to be expected as the
representations that occur as summands are quite restricted, so
that the algebra appears smaller---it is a quotient of Virasoro.
We can see this already for generic $q$, where the standard
modules became the irreducible representations of Virasoro that
lie in the first column of the Kac table. It is possible that the
modules we call projective do obey the general definition of
projective modules over this smaller algebra, even if they do not
for the Virasoro algebra itself. Also, there could in principle be
elements of the algebra not in the Virasoro algebra (though this
seems unlikely in the spin-1/2 chains and related theories studied
here).

When $q$ is a root of unity, the corresponding central charge is
one of the special values of Kac and BPZ (in which $x$ above is
rational), at which the standard modules may acquire additional
null vectors. When this occurs, the standard module of Virasoro
now has the same submodule structure as the standard TL module
illustrated above. The positions in the first column of Kac table
at which this occurs agree exactly with what we would expect based
on the TL algebra. Similarly, we will infer that the projective
and tilting Virasoro modules relevant to the analysis of the XXZ
chain (to operators as well as states, by the state-operator
correspondence) have structures like that above (when reducible).
Finally, the fusion rules must be those inferred from finite size
systems, and for the projective modules are again the same as for
finite size $L$. These can in turn be obtained from those on the
\uqsl2 side, which is very convenient as one can use characters
there. This illustrates how symmetry becomes useful in analyzing
the structure of a CFT.

Thus, in the following sections, the ideas of this part will be
worked out in several instances of physical or geometric interest.


\section{The case $m=0$ ($c=-2$) --- dense polymers}

In this section, we consider the case of the open XXZ chain of
even length $2L$, with $m=0$, so $q=i$ and $r=2$. When the
partition function is expanded as a sum over loops, this model
assigns a factor $m=0$ to each loop, so describes dense polymers.
This case also corresponds (by Morita equivalence of symmetry
algebras) to the gl($n|n$) supersymmetric chains, and to nonlinear
sigma models with target space ${\bf CP}^{n-1|n}$ at a particular
value of the coupling constant, as described in Ref.\
\cite{ReadSal}. For $n=1$, the gl($1|1$) spin chain actually
coincides with the \uqsl2 or spin-1/2 chain, and both are free
fermion systems. In the continuum limit, the latter theory is the
free symplectic fermion theory.

The decomposition of the space of states into a sum of
indecomposable representations (tilting modules) under \uqsl2 has
been well studied. It turns out that none of the indecomposable
summands are simple; all are reducible. Following the notation
described in the previous section, each tilting module ${\rm T}_j$
has the structure corresponding exactly to that in diagram
(\ref{tiltmod}) \cite{PasquierSaleur,Martini,Martinii}:
  $$
  {\rm T}_{j}: \begin{array}{ccccc}
       &&\hskip-.7cmj-1&&\\
       &\hskip-.2cm\swarrow&\searrow&\\
       j-2&&&\hskip-.3cmj\\
       &\hskip-.2cm\searrow&\swarrow&\\
       &&\hskip-.7cmj-1&&
       \end{array},~~~\hbox{$j=1$, \ldots, $L$.}
       $$
The nodes correspond to simple modules and are labeled by spin
values; the simple module of spin $j'\geq 0$ has dimension $j'+1$.
(Nodes with negative spin values are viewed as zero dimensional
and can be omitted.) For some purposes, the module ${\rm T}_0$ for
$j=0$ can also be viewed as a tilting module of \uqsl2 but does
not occur in the chain because its multiplicity is zero (this is
special to the $m=0$ case); in the present section we will view it
also as zero.  Each pair of nodes that can be connected by a
southwest arrow forms a standard module, as in diagram
(\ref{tiltmod}). As advertised, the dimension of the standard
module of spin $j'$ is $2j'+1$; conversely, this together with a
knowledge of the structure enables one to calculate recursively
the dimensions of the simple modules. The total dimension of the
module is thus $4j$. We emphasize that the same structure applies
to the corresponding tilting modules for the enlarged symmetry
algebra ${\cal A})_{n|n}(2L)$, by Morita equivalence, though the
dimensions are different. Similar statements will be true for the
symmetry algebras throughout, but will not usually be made
explicit.

For $L=3$ (i.e.\ six spins $1/2$) for example, the $j$ values of
the indecomposable summands ${\rm T}_j$ are $j=1$, $2$, $3$. The
full decomposition of the chain under \uqsl2 (where $\times n$
denotes multiplicities of the indecomposable tilting modules) is
  \begin{equation}
    \begin{array}{ccc}
      0&\\
      &\hskip-.2cm\searrow\\
      &&\hskip-.3cm1\\
      &\hskip-.2cm\swarrow\\
      0&
      \end{array}
      ~\times 5~~,
     \begin{array}{ccccc}
    &&\hskip-.7cm1&&\\
    &\hskip-.2cm\swarrow&\searrow&\\
    0&&&\hskip-.3cm2\\
    &\hskip-.2cm\searrow&\swarrow&\\
    &&\hskip-.7cm1&&
    \end{array}
    \hskip-.3cm\times 4~~,
    \begin{array}{ccccc}
        &&\hskip-.7cm2&&\\
        &\hskip-.2cm\swarrow&\searrow&\\
        1&&&\hskip-.3cm3\\
        &\hskip-.2cm\searrow&\swarrow&\\
        &&\hskip-.7cm1&&
        \end{array}\hskip-.3cm\times 1\label{uqindec}
    \end{equation}
and the check on the dimensions is
  \begin{equation}
      (1+2+1)\times 5+(2+1+3+2)\times 4+(3+2+3+4)\times 1=64.
      \end{equation}

Knowing the decomposition under (an image of) $U_q($sl$_2)$, one
can also find directly the structure of the modules of the
commutant algebra of \uqsl2 in the spin-1/2 chain. This by
definition consists of all endomorphisms (linear maps of the
vector space into itself) that commute with the \uqsl2 action.
When the algebra is semisimple, as for \uqsl2 at generic $q$, such
endomorphisms can only map a copy of a simple module to another
copy of the same one, and so the commutant is also semisimple, and
has simple modules of dimensions equal to the multiplicities of
the simple modules of the given algebra (this was used in Ref.\
\cite{ReadSal2}). For non-semisimple algebras, as here, there are
such endomorphisms that map the tilting modules to themselves and
which combine to form full matrix algebras. Indeed, the
multiplicities of the \uqsl2 tilting modules are the dimensions of
the simple modules of its commutant. But there are also
endomorphisms that map the indecomposable tilting modules into one
another. This can occur when a tilting module has one or more
subquotient modules that also occur in another tilting module. In
fact the ``top'' of the indecomposable tilting module, which is a
source but not a sink for arrows, must map onto a non-top part of
the other tilting module, and all subquotients that lie below the
top must also map such that the arrows are preserved. In most
cases, the foot of the tilting module, and some subquotients above
it, are mapped to zero. Thus the endomorphisms of the space of the
spin-1/2 chain can be read off from the structure and
multiplicities of the indecomposable tilting modules. For the TL
representation in the spin-1/2 chain, the commutant algebra of
\uqsl2 obtained in this way is isomorphic to the TL algebra; the
two algebras form a ``dual pair''. (If instead the TL structure
were given, one could follow the same procedure to find the
structure of the modules of its commutant.)

 \begin{figure}
 \begin{center}
 \leavevmode
 \epsfysize=80mm{\epsffile{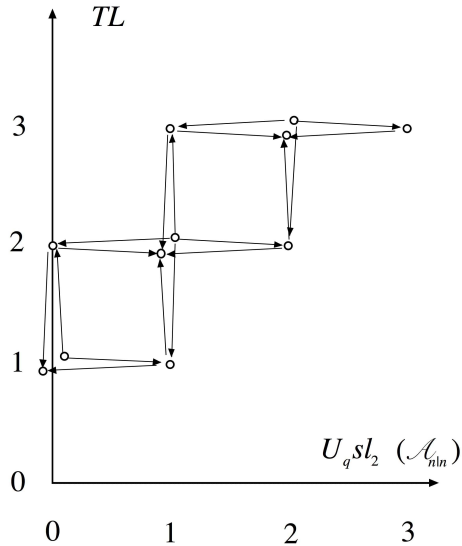}}
  \end{center}
  \protect\caption{The structure of the spin-1/2 chain for $q=i$ and $2L=6$
  sites, as a representation of $U_q($sl$_2)\otimes$TL$_{2L}(q)$, or the
  same for the gl($n|n$)-supersymmetric spin chain with $2L=6$ as a
  representation of ${\cal A}_{n|n}(2L)\otimes$TL$_{2L}(q)$.}
  \label{overall}
 \end{figure}

In the present case, the indecomposable tilting modules ${\rm
T}_j$ have non-trivial homomorphisms into neighboring tilting
modules ${\rm T}_{j\pm 1}$ only. There are also (for $j>0$)
non-trivial homomorphisms of ${\rm T}_j$ into itself, which map
the head to the foot, which have the same $j$ value. The commuting
actions of \uqsl2 and its commutant TL$_{2L}(q)$ can be
represented in a ``staircase'' diagram that generalizes those for
\uqsl2 or TL separately, shown for $L=3$ in Fig.\ \ref{overall}.
This shows the action of both algebras, and thus the tensor
product $U_q($sl$_2)\otimes$TL$_{2L}(q)$, on the states of the
chain. [The diagram is the same if we consider ${\cal
A}_{n|n}(2L)\otimes$TL$_{2L}(q)$ for the supersymmetric chains.]
The circles at Cartesian coordinates $(j,j')$ represent simple
subquotient modules over this product algebra. $U_q($sl$_2)$ acts
by the horizontal arrows (that is, its action preserves $j'$),
while TL$_{2L}(q)$ acts ``vertically'' (that is, preserving $j$),
and these actions commute (readers are invited to check that this
is consistent, because of the arrangement of the arrows as
``commuting squares''). Once again, composites of arrows are
arrows that are implicitly present. Each circle is a subquotient
module with a multiplicity of one when viewed as a
$U_q($sl$_2)\otimes$TL$_{2L}(q)$ module; some values $(j,j')$ do
occur twice, and those circles have been separated slightly for
clarity. The indecomposable tilting modules of \uqsl2 can be
recovered by ignoring vertical arrows, and stretching out the
components of the diagram to recover the diagrams above. Likewise,
the indecomposable tilting modules of the TL algebra are obtained
by ignoring the horizontal arrows. (If we had started knowing
those, then we could have obtained the tilting modules for the
commutant of TL by a similar procedure as we described.) We now
turn to the structure of these in some examples. For this $m=0$,
$L$ integer case, the general form is always a single
indecomposable module under the product algebra, and for $L>3$
simply extends up to larger $j$ values.

In detail for $L=3$, the TL indecomposable tilting modules (direct
summands) have the following structure, in which the nodes are
labelled by the dimension of the corresponding simple subquotient
module over TL, and arrows now denote action of the TL algebra
\cite{martinbook,Martini,Martinii}:
  \begin{equation}
  \begin{array}{ccc}
    (5)&\\
    &\hskip-.2cm\searrow\\
    &&\hskip-.3cm(4)\\
    &\hskip-.2cm\swarrow\\
    (5)&
    \end{array}~\times 1
    ~~,
\begin{array}{ccccc}
      &&\hskip-.7cm(4)&&\\
      &\hskip-.2cm\swarrow&\searrow&\\
      (1)&&&\hskip-.3cm(5)\\
      &\hskip-.2cm\searrow&\swarrow&\\
      &&\hskip-.7cm(4)&&
      \end{array}\hskip-.3cm\times 2~~~,
\begin{array}{ccc}
      &&\hskip-.3cm(1)\\
      &\hskip-.2cm\swarrow&\\
      (4)&&\\
      &\hskip-.2cm\searrow&\\
      &&\hskip-.3cm(1)
      \end{array}\times 3
      ~~ ,(1)\times 4,\label{TLindec}
      \end{equation}
with total dimension
    \begin{equation}
     (5+4+5)\times 1+ (5+4+4+1)\times 2+  (1+4+1)\times 3+ 1\times 4=64
    \end{equation}
Notice here the appearance of multiplicities from the \uqsl2
analysis as dimensions of simple TL modules, and vice versa. Using
the labels $j$ for the TL simple modules as already described,
eq.\ (\ref{TLindec}) becomes
  \begin{equation}
  \begin{array}{ccc}
    1&\\
    &\hskip-.2cm\searrow\\
    &&\hskip-.3cm2\\
    &\hskip-.2cm\swarrow\\
    1&
    \end{array}~\times 1
    ~~,
\begin{array}{ccccc}
      &&\hskip-.7cm2&&\\
      &\hskip-.2cm\swarrow&\searrow&\\
      1&&&\hskip-.3cm3\\
      &\hskip-.2cm\searrow&\swarrow&\\
      &&\hskip-.7cm2&&
      \end{array}\hskip-.3cm\times 2~~~,~~~
      \begin{array}{ccc}
        &&\hskip-.3cm 3\\
        &\hskip-.2cm\swarrow&\\
        \hskip-.3cm 2&&\\
        &\hskip-.2cm\searrow&\\
        &&\hskip-.3cm 3
        \end{array}~\times 3,~~ 3\times 4,\label{TLindeci}
      \end{equation}
For general $L>1$, the pattern is similar, starting with a
``trio'' at $j=1$, followed by quartets up to $j=L$, and then a
single simple module also at $j=L$ (here we assign $j$ to the TL
tilting modules by the $j$ of the head).

The tilting modules for TL can also be compared with the standard
modules. The standard module labeled $j$ for TL$_{2L}(q=i)$
consists of the simple modules $j$, $j+1$, together with the TL
action that connects them by an arrow, except for $j=L$ when the
simple module is standard and one dimensional ($d_L=1$), and $j=0$
where the standard module is simple. For six spins $1/2$ again,
one has $d_{0}=5$, $d_{1}=9$, $d_{2}=5$, $d_{3}=1$. The
indecomposable tilting modules consist of pairs of standard
modules ``glued'' together by the action of some TL elements,
similarly to $U_q($sl$_2)$.

When the structure of the tilting modules is known as in
(\ref{TLindeci}), one can recursively calculate the dimensions
$d_j^0$ of the simple modules of TL, starting from $j=L$. For
$L=3$ this gives $d_3^0=1$, $d_2^0=d_2-d_3=4$,
$d_1^0=d_1-d_2^0=5$, $d_0^0=d_0-d_1^0=0$. Usually there is a
simple module of TL for $j=0$, but here for $m=0$ it vanishes.
(This is related with the fact that the partition function of the
chain vanishes exactly). For general even $2L$, one finds the
formula for the dimensions of simple TL modules for the $q=i$ case,%
\be%
d_j^0=\sum_{j'=j,j+1,\ldots,L}(-1)^{j-j'}d_{j'}\ee%
for $j=0$, $1$, \ldots, $L$.

 %
     %

We may also consider $U_q($sl$_2)^{(2L)}$ and analyze it viewed as
a module over itself, acting on the left or on the right (the left
and right regular representations), or on both sides (i.e.\ as a
$U_q($sl$_2)^{(2L)}\otimes U_q($sl$_2)^{(2L)}$ module, as the
actions on the two sides commute). From the last of these, we can
then make a similar staircase diagram for $U_q($sl$_2)^{(2L)}$ as
a module over $U_q($sl$_2)^{(2L)}\otimes U_q($sl$_2)^{(2L)}$, from
which the left or right structure can be determined. The cellular
ideals we reviewed in Ref.\ \cite{ReadSal2} can be readily seen in
such a picture, as they form sets of circles and arrows mapped
into themselves by the left and right action of
$U_q($sl$_2)^{(2L)}$. This is also useful in calculating the total
dimension of the algebra, which is equal to ${\rm dim}\,
U_q($sl$_2)^{(2L)}=\sum_{j=0}^L (2j+1)^2$, even though it is not
semisimple. One can do the same for the TL algebra, and then ${\rm
dim}\, {\rm TL}_{2L}(q)=\sum_{j=0}^Ld_j^2$ ($=132$ for $L=3$).
Similarly, for the spin-1/2 chain itself, we can point out that
the total dimension is $2^{2L}=\sum_{j=0}^L (2j+1)d_j$, again just
as if the algebras were semisimple. Of course all of these
expressions for the dimensions follow from the independence of the
dimensions of the standard modules from $q$, so that the formulas
from the generic case always hold. This is also true for the other
non-generic values of $q$, some of which are considered later. For
the models with Morita equivalent symmetry algebras, $2j+1$ should
be replaced by the dimensions $D_j'=[2j+1]_{q'}$, where
$q'+q'^{-1}=2n$ \cite{ReadSal2}, but again it is true that the
formulas can be obtained ``as if'' the standard modules were
simple (similar formulas also hold for the $q$-dimensions [in the
\uqsl2 case] and the superdimensions [in the superalgebra cases
\cite{ReadSal2}]).

We will give here the expressions for the (super-)dimensions of
the simple modules of our algebras ${\cal A}_{n|n}$, $n=1$, $2$,
\ldots, which are Morita equivalent to \uqsl2 with $q=i$
\cite{ReadSal2}. First we point out that the dimensions $j+1$ of
the $j$th simple module of \uqsl2 can be deduced from the
structure of the standard modules, and the dimension $2j+1$ of the
latter, much like those for TL. They are obtained from the
relations $j+1=(2j+1)-j$, with dimension $1$ for the simple module
with $j=0$, which follow from the existence of simple sub- and
quotient modules of the each standard module. For the algebras in
the supersymmetric models, the standard modules have dimensions
$D_j'=[2j+1]_{q'}$, where $q'+q'^{-1}=2n$ \cite{ReadSal2}. Then as
the structure of submodules and quotients is the same as for
\uqsl2 because of the Morita equivalence \cite{ReadSal2}, we have
for the dimensions $D_j'^0$ of the simple modules the
corresponding recurrence $D_j'^0=D_j'-D_{j-1}'^0$, with
$D_0'=D_0'^0=1$ obviously corresponding to a simple module. This
is solved by
\bea
D_j'^0&=&D_j'-D_{j-1}'+D_{j-2}'-\ldots\nonumber\\
&=&q'^{2j}+q'^{2j-4}+\ldots+q'^{-2j}\nonumber\\
&=&[j+1]_{q'^2}.
\eea
Similarly, we may look at the quantum dimensions of the modules;
for the algebra ${\cal A}_{n|n}$ the quantum dimensions are just
the superdimensions. For the $j$th standard module, the
superdimensions were denoted $D_j$ in Ref.\ \cite{ReadSal2}, given
by $D_j=[2j+1]_q$, which is the same as the $q$-dimension for
\uqsl2 because of the Morita equivalence with \uqsl2 as ribbon
Hopf algebras (here for the algebras with integer $j$ only)
\cite{ReadSal2}. In the present case, $D_j=(-1)^j$. The same
equality applies to dimensions of other modules such as the simple
modules, and thus we can compute the quantum dimensions for \uqsl2
($q=i$) simples and ${\cal A}_{n|n}$ simples as
$D_j^0=[j+1]_{q^2}$ similarly. For the tilting modules,  the
quantum dimensions are $D_j^+=D_j+D_{j-1}$ ($j=1$, $2$, \ldots),
and so $D_j^+=0$ for all $j\geq 1$.

For a finite-dimensional algebra, the indecomposable summands in
the left regular representation are the indecomposable projective
(left) modules. Thus we can read off the indecomposable projective
modules. For the TL algebra (for which these are of particular
interest) for $L=3$, these are the same modules as the three
tilting modules other than the singlet. The singlet cannot be an
indecomposable projective module as it occurs as a quotient module
of an indecomposable tilting module, which is impossible for a
projective module (one of the alternative definitions for a
projective module is that whenever it is a quotient of some
module, it is isomorphic to a direct summand in that other
module). The other three indecomposable tilting modules are
acceptable, and are projective as follows automatically if the TL
algebra is defined as the commutant of \uqsl2 in this chain (this
is an aspect of the key lemma \cite{lemma}; see also Ref.\
\cite{ReadSal2}). There are no other indecomposable projective
modules in the present case (for the other cases $m\neq0$, we will
see that there are also projective modules that are not tilting).
So the indecomposable projectives of the TL algebra are
\begin{equation}
 {\rm P}_j: \begin{array}{ccc}
    1&\\
    &\hskip-.2cm\searrow\\
    &&\hskip-.3cm2\\
    &\hskip-.2cm\swarrow\\
    1&
    \end{array}~\times 1
    ~~,
\begin{array}{ccccc}
      &&\hskip-.7cm2&&\\
      &\hskip-.2cm\swarrow&\searrow&\\
      1&&&\hskip-.3cm3\\
      &\hskip-.2cm\searrow&\swarrow&\\
      &&\hskip-.7cm2&&
      \end{array}\hskip-.3cm\times 2~~~,
      ~~\begin{array}{ccc}
        &&\hskip-.3cm 3\\
        &\hskip-.2cm\swarrow&\\
        \hskip-.3cm 2&&\\
        &\hskip-.2cm\searrow&\\
        &&\hskip-.3cm 3
        \end{array}~\times 3,
      \end{equation}
This pattern is repeated for general $L>0$ (integer); the
indecomposable projectives ${\rm P}_j$ of TL for $q=i$ are the
indecomposable tilting modules, other than the singlet at $j=L$: %
\be%
{\rm P}_j:\begin{array}{ccccc}
       &&\hskip-.7cm j&&\\
       &\hskip-.2cm\swarrow&\searrow&\\
       j-1&&&\hskip-.3cm j+1\\
       &\hskip-.2cm\searrow&\swarrow&\\
       &&\hskip-.7cm j&&
       \end{array},~~~\hbox{$j=1$, \ldots, $L$,}
\ee%
where once again we understand in this case that the simple
modules with $j<1$ or $j>L$ are zero and can be omitted. (For
$L=1$, one does have a two-dimensional indecomposable ${\rm P}_1$,
because of the implicit arrow from the head $j=1$ to the foot
$j=1$.)

\begin{figure}
  \begin{center}
   \leavevmode
   \epsfysize=80mm{\epsffile{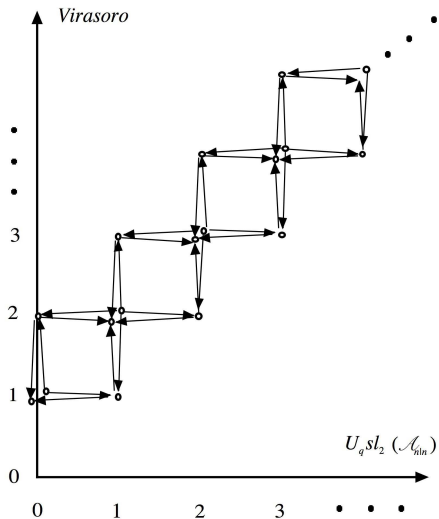}}
   \end{center}
   \protect\caption{Space of states of the continuum $m=0$ theory,
   showing the commuting actions of Virasoro and $U_q($sl$_2)$ (or
   ${\cal A}_{n|n}$).}
  \label{contoverall}
  \end{figure}

We now pass to the continuum limit. Our strategy will be to deduce
as much as possible from the lattice, and compare with known
results from the representation theory of the Virasoro algebra.
First, we can consider the purely algebraic limit $L\to\infty$ as
discussed in Ref.\ \cite{ReadSal2}. The staircase diagram then
becomes infinite; see Fig.\ \ref{contoverall}, and the limits of
the dimensions $d_j^0$ of the simple TL modules are also infinite.
The more physical way of taking $L\to\infty$, called the continuum
limit \cite{ReadSal2}, focusses on low-energy states and the
long-wavelength Fourier components of the $e_i$s. We expect that
the Virasoro algebra emerges from the TL algebra in this limit
\cite{KooSal}, and that the staircase diagram still applies to
this. In addition we will use the known information (as for the
case of generic $q$) of the values of the central charge and a set
of scaling dimensions that are associated with the Virasoro
modules (or the scaling fields of the theory). This is done for
convenience; we expect that with some additional effort these may
themselves be established purely on the grounds of consistency
with the algebraic structure encoded in the staircase diagram,
since the structure of the \uqsl2 modules is known and is
unchanged in the $L\to\infty$ limit.

On specializing the generic results, eqs.\ (\ref{c}), (\ref{kac})
for the central charge and the Kac table to $q=i$, so $x=1$, one
finds that $c=-2$, and the Kac formula reads
\begin{equation}
    h_{rs}=\frac{(2r-s)^{2}-1}{8},
    \end{equation}
where $r$, $s$ are positive integers. The values appearing in the
spectrum are
\begin{equation}
    h_{1,1+2j}=h_{j,1}=\frac{(2j-1)^{2}-1}{ 8}=\frac{j(j-1)}{2},
    \end{equation}
where we used symmetries of the Kac formula to restrict to the
fundamental domain, and again here $j\geq 0$ is integer.

As the limits of the dimensions of the modules over TL are
infinite, it is convenient to use instead the graded dimensions or
characters defined using a trace over the module of the
exponential of the Hamiltonian, or $L_0$ Virasoro generator. For
the standard modules, which we will denote $r_{r,s}$, these were
given in Sec.\ 2, and for the cases of interest here are
\begin{equation}
    K_{1,1+2j}={\rm Tr}_{r_{1,1+2j}}\, \widehat{q}^{~L_0-c/24}=
    \frac{\widehat{q}^{~(2j-1)^{2}/8}-\widehat{q}^{~(2j+3)^{2}/8}}{
    \eta(\widehat{q})}.
    \end{equation}
We may point out here that the full partition function of the
model in the continuum limit is obtained by specializing the result for
generic $q$: %
\be%
Z={\rm Tr}\, \widehat{q}^{~L_0-c/24}=
\sum_{j=0}^\infty(2j+1)\frac{\widehat{q}^{~(2j-1)^{2}/8}-
\widehat{q}^{~(2j+3)^{2}/8}}{\eta(\widehat{q})},\ee%
which as for the total dimension of the finite length chains is
still valid at $q=i$, even though the modules involved are no
longer fully decomposable. This is similar to the result given for
the case $c=0$ in the appendix of Ref.\ \cite{ReadSal}. This
result may be obtained using field-theoretic methods \cite{sb},
and provides a check on our algebraic deductions. It is perhaps
more natural to calculate a partition function as a quantum trace
instead of a trace, which corresponds to a boundary condition in
the (periodic, imaginary) time direction that does not break the
\uqsl2 or ${\cal A}_{n|n}$ symmetry. In that case $2j+1$ is
replaced by $D_j$, and the result is valid for the continuum limit
of the supersymmetric models as well as that of the spin-1/2 chain
\cite{ReadSal}. In the present case $q=i$, these partition
functions (called ``unmodified'' in Ref.\ \cite{ReadSal}) vanish
identically (this is also true on the lattice). This follows
because we can collect all states into tilting modules of the
symmetry algebra,  and the quantum dimensions $D_j^+$ of these
tilting modules vanish. For the modified partition functions of
the supersymmetric models, $2j+1$ must be replaced by $D_j'$ as
for the finite-$L$ chains.

\begin{figure}
   \begin{center}
 \leavevmode
 \epsfysize=10mm{\epsffile{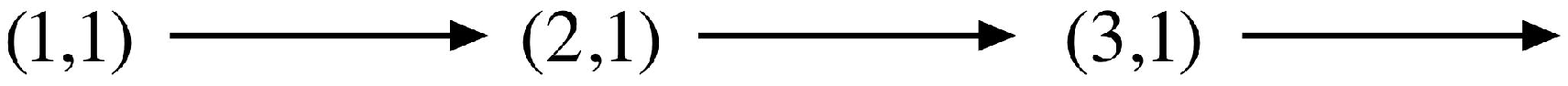}}
 \end{center}
 \protect\caption{Chain of Verma modules embeddings for $c=-2$. }
   \label{vermemb}
\end{figure}

In studying the representation theory for highest-weight modules
of the Virasoro algebra, it is conventional to use Verma modules.
The Verma module generated from the highest weight vector with
weight $h_{1,1+2j}$ has, for generic central charge, a singular
vector at level $1+2j$, with value of the conformal weight
$h_{1,-1-2j}$. That is, there is a submodule which is also highest
weight with that conformal weight, and vectors in the submodule
are orthogonal to all those in the Verma module. At $c=-2$,
another singular vector appears at the lower level
$h_{j+1,1}=h_{j,1}+j$, while now
$h_{1,-1-2j}=h_{j+2,1}=h_{j,1}+1+2j$. These submodules in turn
contain submodules. This leads to a picture of Verma modules
embedded into one another in a sequence, as shown in Fig.\
(\ref{vermemb}).

The representations obtained by factoring out from the Verma
module the submodule generated from the ``generic'' singular
vector at level $1+2j$ are the standard modules $r_{1,1+2j}$. At
$c=-2$ this representation is reducible; from Fig.\ \ref{vermemb}
it contains a submodule that is simple, and its structure is
described by a diagram of the form (\ref{stanmod}). It possesses
an irreducible quotient $R_{j,1}$, with character $\chi_{j,1}$,
and the character of the submodule in $r_{1,1+2j}$ is
$\chi_{j+1,1}$. Hence we have the equality
 \begin{equation}
     K_{1,1+2j}=\chi_{j,1}+\chi_{j+1,1}.
     \end{equation}
The calculation of $\chi_{j,1}$ from the embedding diagram Fig.\
\ref{vermemb} is standard, and gives
 \begin{equation}
     \hbox{Tr }_{ R_{j,1}}\widehat{q}^{~L_{0}-c/24}=\chi_{j,1}=
     \frac{\widehat{q}^{~(2j-1)^{2}/8}-\widehat{q}^{~(2j+1)^{2}/8}}{
     \eta(\widehat{q})},
     \end{equation}
and note that $\chi_{0,1}=0$ (the continuum equivalent of the
absence of the $(0,0)$ node in Fig.\ \ref{overall}).

This shows that the structure of the standard Virasoro modules at
$c=-2$ parallels that of the finite-$L$ TL modules of the same
$j$. We now go a step further and describe the structure of
Virasoro modules that parallel the indecomposable projective
modules of TL; these will be called indecomposable projective
modules of Virasoro (the terminology is subject to caveats already
mentioned in the previous section). The modules we want are the
continuum limits of the finite-$L$ indecomposable projective TL
modules. Calling these indecomposable projective modules ${\cal
R}_{j}$, we expect thus in particular to have the characters
 \begin{equation}
     \hbox{Tr
     }_{{\cal R}_{j}}\widehat{q}^{~L_{0}-c/24}=K_{1,1+2j}+K_{1,1+2(j-1)}=
     2\chi_{j,1}+\chi_{j-1,1}+\chi_{j+1,1}.
     \end{equation}

\begin{figure}
 \begin{center}
  \leavevmode
  \epsfysize=40mm{\epsffile{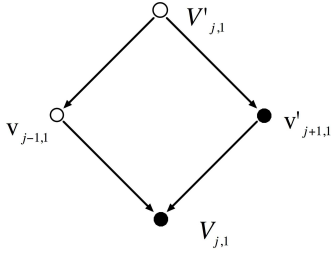}}
  \end{center}
  \protect\caption{The submodule structure of the indecomposable
  projective modules of the Virasoro algebra. The vectors are
  highest weights
  in the corresponding simple subquotient modules. Open dots mean the
  corresponding vectors are not Virasoro descendants of any other
  state within the representation.}
 \label{repstruc}
 \end{figure}

The structure of these modules, in the sense of the submodules
that appear, is as follows. All are described by the same diagrams
as the corresponding projectives ${\rm P}_j$ of TL, and all except
that for $j=1$ are ``quartets'' since the effects at $j=L$ have
disappeared as $L\to\infty$. Each circle in the diagram is a
simple subquotient module, and for the Virasoro algebra these are
highest weight modules. In more detail, we should first have an
irreducible representation $R_{j,1}$ generated by a highest weight
vector $V_{j,1}$ at the ``foot''; then a standard submodule
$r_{j-1,1}$ generated by a highest weight vector $v_{j-1,1}$, in
which $V_{j,1}$ appears as a singular vector. Then a state
$V'_{j,1}$ which forms a two dimensional Jordan cell (of $L_{0}$)
with $V_{j,1}$:
\begin{eqnarray}
   L_{0}V'_{j,1}&=&h_{j,1}V'_{j,1}+V_{j,1},\nonumber\\
  L_{0} V_{j,1}&=&h_{j,1}V_{j,1}.
 \end{eqnarray}
$V'_{j,1}$ is not a highest weight vector, and positive modes of
Virasoro (i.e.\ $L_n$'s with $n>0$) map it to descendants of
$V_{j-1,1}$. Finally, there should be a  $v'_{j+1,1}$ which is
technically a sub-singular vector, i.e. it is singular in the
quotient by the representation generated by $V_{j,1}$ (in other
words, positive modes of Virasoro send it to descendants of
$V_{j,1}$). This is schematically represented in Fig.\
\ref{repstruc}, where arrows indicate action of the Virasoro
algebra. Nodes connected by southeast arrows contribute
$K_{1,1+2j}$ and $K_{1,1+2(j-1)}$ to the character; these
correspond to a submodule and a quotient module, both of which are
isomorphic to standard modules; the head and foot of each of these
are indicated by the open and closed circles. In the CFT, these
vectors that are highest weights modulo submodules will be the
closest that we can come to finding primary fields for the
Virasoro algebra.

Representations with the quartet structure such as ${\cal R}_{j}$
have been abstractly studied before and are called staggered
modules in \cite{Rho}. It can be shown that they depend on one
characteristic parameter, which can be defined through the
equation
 \begin{eqnarray}
    L_{1}^{j-1}V'_{j,1}=\beta_{j}v_{j-1,1},\nonumber\\
    L_{k}V'_{j,1}=0,~~~k\geq 2.
    \end{eqnarray}
We do not know the relevant values of $\beta_{j}$ in our models;
it would be interesting to obtain them from the spin-1/2 chain, or
from the symplectic fermion field theory discussed below.
Presumably our modules are special as they are ``projective.''

Now we interpret the continuum limit using the symplectic fermion
representation of the theory. This follows from writing the
spin-1/2 chain, or the supersymmetric formulation for $n=1$, as a
free-fermion system. The continuum limit can then be taken. As we
consider the open chain case, one must treat the boundary
conditions on the fermion fields. The fermion fields can be viewed
as a complex scalar fermion, and its conjugate. These fermions are
partners under a supersymmetry (graded Lie algebra) of free
bosons, and the corresponding continuum free boson theory is a
complex scalar boson. The appropriate boundary conditions for the
latter were considered in detail in Ref.\ \cite{xrs}, both for the
lattice model (discrete in both space and time directions, and
called a network model) and in the continuum. Hence the boundary
condition on the fermions also is a variant of a Neumann boundary
condition, in which on the boundary, the derivative of the field
at some fixed angle (the ``Hall angle'' \cite{xrs}) to the normal
to the boundary must vanish (the Hall angle is a free parameter of
the theory). This is due to the alternation of the two types of
site, in the fundamental and its dual, along the chain when viewed
as a representation of gl($1|1$). A property of these boundary
conditions is that, as for ordinary Neumann boundary conditions,
when we find the normal modes of the continuum theory on a finite
interval, there is a zero mode for the fermion. Indeed, the entire
spectrum is independent of the choice of Hall angle. For any value
of the Hall angle, the theory can be mapped to the Hall angle zero
case, which is the ordinary Neumann boundary condition. The
transformation is a dual rotation, mapping the field to a linear
combination of itself and its dual field; this transformation is
nonlocal in terms of the fields, but linear, and commutes with the
U(1) symmetry of the complex fermion theory. After this
transformation, the fields of the theory are the fermion fields
$\psi$ with two real components $\psi_1$, $\psi_2$, where
$\psi=\psi_1+i\psi_2$, and their derivatives with respect to $z$
only (as they are boundary fields), and products of these. The
Lagrangian density is
$\frac{1}{2}(\partial_\mu\psi_1\partial_\mu\psi_2-\partial_\mu
\psi_2\partial_\mu\psi_1)$, and thanks to the Neumann boundary
condition, there is a well-known exact symplectic [Sp(2,{\bf R})]
symmetry of this theory, that acts locally on the fields. Note
however that for the original field variables that were natural
for the spin chain, only the U(1) subgroup of this symmetry acts
locally (it corresponds to $S^z$ or fermion number symmetry in the
chain), while the other two generators of the sp$_2$ Lie algebra
are nonlocal in those variables and arise as the ``renormalized
squares'' of $S^+$ and $S^-$ in the \uqsl2 language, or as some of
the nonlocal $k=2$ basis elements constructed in Ref.\
\cite{ReadSal2}.

Now we classify these local operators according to the Virasoro
[and $U_q($sl$_2)$] module structure we have worked out. The
arrows connecting simple modules in the indecomposables of either
algebra are always due to the fermion zero modes \cite{Kausch1},
which correspond to the local fields $\psi_i$ without derivatives,
as in operator products $\partial \psi_1(z)\psi_2(0)\sim 1/z$,
while $\psi_i$ without a derivative cannot be produced by any
operator products of the local fields if it is not already a
factor in an operator. The case $j=2$ is typical. Then
$h_{j,1}=1$, the vector $V_{j,1}$ can be identified e.g.\ as the
derivative of either symplectic fermion, say
$V_{j,1}=\partial\psi_{2}$. The vector $v_{1,1}=\psi_{2}$ has
dimension $h_{1,1}=0$ and is a Virasoro highest weight: it
generates a reducible representation, in which $\partial\psi_{2}$
appears as a singular vector. We still need a field $v'_{3,1}$
which is a combination involving
$\psi_{1}\partial\psi_{2}\partial^{2}\psi_{2}$ etc. It is not
primary, nor is $v'_{2,1}=\psi_{1}\psi_{2}\partial\psi_{2}$. By
interchange $\psi_{1}\leftrightarrow\psi_{2}$, two such Virasoro
modules appear, which corresponds to the dimension of the \uqsl2
simple module, which is just 2.

\begin{figure}
 \begin{center}
  \leavevmode
  \epsfysize=40mm{\epsffile{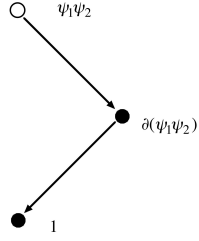}}
  \end{center}
  \protect\caption{Indecomposable representation for $j=1$.}
 \label{partindec}
 \end{figure}

For the case $j=1$, the cell lacks its left hand side and looks as
on  figure \ref{partindec}. One has for instance
$L_{-1}(\psi_{1}\psi_{2})=\partial(\psi_{1}\psi_{2})$ while
$L_{1}(\partial(\psi_{1}\psi_{2}))=1$, etc.

For the \uqsl2 structure, we may notice that general local fields
that are highest weights for Virasoro, modulo submodules, can be
written as one of the following form: (i) ${\cal
S}(\partial\psi_{i_1}\partial^2\psi_{i_2}
\cdots\partial^{j-1}\psi_{i_{j-1}})$, with $j=1$, $2$, \ldots,
where the symmetrizer $\cal S$ symmetrizes in the indices $i=1$,
$2$; (ii) the same with $\psi_{i_0}$ inserted inside the
symmetrizer $\cal S$; (iii) a similar expressions as in (ii) but
with the index on $\psi_{i_0}$ contracted with one of the
$\partial^k\psi_i$ using the symplectic form; (iv) the form (i)
times $\psi_1\psi_2$. For each $j$, these all have the same
scaling dimension (meaning the eigenvalue of $L_0$, modulo
submodules). Those of form (i) are the foot of the Virasoro module
${\cal R}_j$, that is they form a simple module of scaling
dimension $j(j-1)/2$; this may be seen using operator products
with the stress tensor
$T=-\frac{1}{2}(\partial\psi_1\partial\psi_2
-\partial\psi_2\partial\psi_1)$. There are $j$ distinct such
operators, because of the symmetrization, and so they form a
simple \uqsl2 module, with the \uqsl2 spin $j-1$. The local fields
of form (ii) in which $\psi_{i_0}$ are included form a multiplet
of dimension $j+1$, those of type (iii) form a multiplet of
dimension $j-1$, and finally those of form (iv) in which
$\psi_1\psi_2$ is included form a multiplet of dimension $j$. For
each $j$, these four types of local operators (or three types if
$j=1$) make up the tilting module ${\rm T}_{j}$ of $U_q($sl$_2)$,
listed here starting with the foot and ending with the head. The
symplectic or sp$_2$ symmetry of symplectic fermions acts to map
the fields of the same type into each other; if we view the
boundary symplectic fermion theory as a chiral theory in the
plane, then there are three corresponding Noether currents ${\cal
S}(\psi_i\partial\psi_{i'})$. The doublet of currents
$\partial\psi_i$ of psl($1|1$) \cite{ReadSal} act (through
operator product) to remove an undifferentiated $\psi_i$ from the
boundary local fields. These are the arrows in the \uqsl2
quartets. The \uqsl2 algebra can here be viewed as isomorphic to a
Lie superalgebra generated by the line integrals of these five
local currents (which also correspond to the five generating
elements of \uqsl2 at a root of unity); this Lie superalgebra is
not semisimple (as a Lie superalgebra), because the psl($1|1$)
generators form an ideal.

For the Virasoro structure, it suffices to notice that the stress
tensor $T$ also maps each of the four types into themselves by
operator products, and also can reduce the number of
undifferentiated fermion fields in a local operator by one or two.
These actions then exhibit the commuting actions of Virasoro and
$U_q($sl$_2)$, and show how the staircase diagram structure
appears in terms of the local fields.

The partition function can also be interpreted in terms of
symplectic fermions. In terms of the characters of simple Virasoro
modules and dimensions of simple modules of $U_q($sl$_2)$, it
becomes
\begin{equation}
    Z=\sum_{j=0}^{\infty}j(2\chi_{j,1}+\chi_{j-1,1}+\chi_{j+1,1})
    =
    \frac{2\theta_{2}(\widehat{q})}{\eta(\widehat{q})}
    =\hbox{det}\left(-{\cal D}_{A,N}\right),
    \end{equation}
where ${\cal D}_{A,N}$ is the Laplacian on the strip with Neumann
boundary conditions in the space direction and antiperiodic
boundary conditions in the time direction. For symplectic
fermions, the vanishing of the ``unmodifed'' (or quantum trace)
partition function can be viewed as due to the fermion zero modes,
as is well known (in the present language, it is the determinant
of the Laplacian with periodic boundary conditions in the time
direction, and this Laplacian has a zero mode).

We now discuss fusion of the tilting modules ${\rm T}_{j}$ (of
dimension $4j$) labelled by $j$, in other words the Clebsch-Gordan
series for \uqsl2 at $q=i$. A tensor product of indecomposable
tilting modules necessarily decomposes into a direct sum of
tilting modules, and given this fact, the decomposition can be
easily determined by using characters \cite{PasquierSaleur}. It
takes the form:
\begin{equation}
    {\rm T}_{j_{1}}\otimes {\rm T}_{j_{2}}={\rm T}_{j_{1}+j_{2}}\oplus
    2 {\rm T}_{j_{1}+j_{2}-1}\oplus\ldots\oplus 2{\rm
    T}_{|j_{1}-j_{2}|+1}\oplus {\rm
    T}_{|j_{1}-j_{2}|}.\label{firstfus}
    \end{equation}
Here $2T$ means $T\oplus T$, and for $j_1=j_2$, the term $T_0$ is
absent; for the present we may view this exception as handled by
our convention that ${\rm T}_j=0$ if $j<1$. (A check on the result
can be easily obtained by considering the dimensions $2j+1+(2j-1)$
of the tilting modules.) It follows from Morita equivalence that
the same fusion rules apply to the tilting modules of our
supersymmetry algebras ${\cal A}_{n|n}$ also \cite{ReadSal2}.

From the general theory \cite{ReadSal2}, the fusion rules for the
product {\em functor} for the indecomposable {\em projective} TL
modules have the same form, %
\begin{equation}
    {\rm P}_{j_{1}}\times {\rm P}_{j_{2}}={\rm P}_{j_{1}+j_{2}}+
    2\!\sum_{j=|j_1-j_2|+1}^{j_1+j_2-1} {\rm P}_j+ {\rm
    P}_{|j_{1}-j_{2}|},
    \end{equation}
and so do those for the projective Virasoro modules in the
continuum limit:
\begin{equation}
    {\cal R}_{j_{1}}\times {\cal R}_{j_{2}}={\cal R}_{j_{1}+j_{2}}+
    2\!\sum_{j=|j_1-j_2|+1}^{j_1+j_2-1} {\cal R}_j+ {\cal
    R}_{|j_{1}-j_{2}|}.
    \end{equation}
(Again, recall that the modules ${\rm P}_0$ and ${\cal R}_0$ are
zero.) The summations may be re-organized into the form
\begin{equation}
  {\cal R}_{j_{1}}\times {\cal R}_{j_{2}}={\sum_{j}}'(2{\cal
   R}_{j}+{\cal R}_{j-1}+{\cal R}_{j+1}),\label{secondfus}
   \end{equation}
where the sum is over $j=j_{1}+j_{2}-1$, $j_{1}+j_{2}-3$, \ldots,
$|j_{1}-j_{2}|+1$. This shows that these results agree with those
for $c=-2$ in Ref.\ \cite{Gaberdielindec}. There, some
indecomposables of Virasoro are built somewhat explicitly, and
fusion is defined and calculated up to level six through the
algorithm in \cite{Gaberdielfusion}; the final results are then
conjectured. The correspondence in the latter reference would be
$m\equiv j$ and ${\cal R}_{m,1}\equiv {\cal R}_{j}$. Calculations
in \cite{Gaberdielindec} are extremely technical, and it is quite
remarkable that the results can be derived from fusion of
indecomposables of $U_{q}($sl$_{2})$.

%
    %
%

%

We have concentrated here on the fusion rules for tilting modules
on the \uqsl2 side, and the projective modules on the TL or
Virasoro side, as these are the most natural starting point, and
there is an equivalence between these fusion products. One can
also derive from these the fusion rules for some other types of
modules, such as the standard and simple modules. One must be
careful in using the \uqsl2 fusion results on the TL/Virasoro side
as the equivalence may not always hold for the non-tilting
modules. The general definition of fusion for the TL or Virasoro
modules is always the module induced from the tensor product by
the inclusion of the algebras for two chains into that for a
single chain, by joining them end to end.

We emphasize that these fusion rules can be viewed as describing
operator products of the conformal fields that transform as
Virasoro modules discussed here, for example for the symplectic
fermions, for which the ``pseudoprimary'' fields were written down
earlier, each of which is the primary field in a simple
subquotient module. They also apply to the corresponding fields in
the supersymmetric models, where the fields have different
multiplicities given by the dimensions $D_j'^0$. In applications
the fusion rules given here will have to be combined with those
for the symmetry algebra [\uqsl2 or ${\cal A}_{n|n}$)], as the
fields transform under both algebras. These combined forms can be
obtained in the finite-$L$ lattice models, where they come from
decomposing tensor products simply as vector spaces (the
multiplicities from the symmetry algebra side make the dimensions
of spaces conserved), and then the continuum limit can be taken.

As a final note, we mention that in the case of unoriented loops,
we should consider odd as well as even values of $2L$. Then the
half integer spins $j$, for which $2j$ is odd also appear in the
spectrum. For $q=i$, these standard modules remain simple, both
for \uqsl2 and (hence) also for TL. They appear then as extra
circles interspersed among the integer $j$ values in the staircase
diagrams, with no extra explicit arrows, so these modules are both
tilting and projective also. Fusion rules for products involving
these can be found along the same lines as for $j_1$, $j_2$
integer (the fusion rules for a product of $j_1$ with $j_2$ both
integers are unchanged, and still close on integer $j$ only). In
terms of dense polymers, the unoriented-loops model contains local
boundary fields for odd as well as even numbers of legs, the
number of legs being (as always for the loop models) $2j$ for
fields in a standard module of spin $j$.


\section{The case $m=1$ ($c=0$) --- percolation}

In this section we follow a similar procedure as in the previous
one, for the XXZ chain of even length $2L$ with $m=1$, so
$q=e^{i\pi/3}$. In this case the loop model describes dense,
oriented loops which can be identified with the boundaries of
clusters in percolation at the critical percolation threshold. In
the spin chains with gl($n+1|n$) supersymmetry \cite{ReadSal}, the
continuum limit in this case corresponds to the fixed-point theory
of the ${\bf CP}^{n|n}$ nonlinear sigma model at $\theta=\pi$,
with the enlarged symmetry algebra ${\cal A}_{n+1|n}$ that is
Morita equivalent to \uqsl2 with $q=e^{i\pi/3}$ \cite{ReadSal2}.
The $n=1$ case is thus connected with the spin quantum Hall
transition also \cite{glr}. Because the TL algebra for $-m$ is
isomorphic to that for $m$, the first, algebraic part of the
analysis is the same for the $m=-1$ finite-$L$ spin chain for that
case also, up to a change in sign of the Hamiltonian. Because of
the latter sign change, a distinct continuum limit is obtained for
$m=-1$ which is not considered in this section. We note that since
the chain has even length, only integer spins $j$ occur, and for
these the symmetry algebras \uqsl2 and $U_{-q}($sl$_2)$ are
isomorphic.

We start with the analysis of the spin-1/2 chain under
$U_{q}($sl$_{2})$. For $q=e^{i\pi/3}$, so $r=3$, and the
representations depend on $2j+1$ modulo $3$, instead of modulo $2$
as in the previous case. The standard modules for $j\geq 0$ with
$j\equiv 1$ (mod 3) remain irreducible. The remaining standard
modules for $j>0$ become reducible in the manner described for all
roots of unity, and the corresponding tilting modules (which are
the direct summands in the spin-1/2 chain) contain a submodule and
a quotient module that are standard, and which have adjacent
$j$-values if the values $j\equiv 1$ (mod 3) are skipped.
Similarly, the $j$ values assigned to the simple subquotient
modules of the standard modules are also adjacent if the values
$j\equiv1$ (mod 3) are skipped, and the value at the head is the
same as the value for the standard module. The standard module for
$j=0$ is one dimensional, and so is simple, and also appears as a
summand in the chain. Thus the tilting modules have the following
structure \cite{PasquierSaleur,Martini,Martinii}:
\be%
{\rm T}_{j}:~~~~~\left\{\begin{array}{cl}
\begin{array}{ccccc}
     &&\hskip-.7cmj-1&&\\
     &\hskip-.2cm\swarrow&\searrow&\\
     j-3&&&\hskip-.3cmj\\
     &\hskip-.2cm\searrow&\swarrow&\\
     &&\hskip-.7cmj-1&&
     \end{array}&\hbox{$j\equiv0$ (mod 3),}\nonumber\\
     &\nonumber\\
      j&\hbox{$j\equiv1$ (mod 3),}\nonumber\\
      &\nonumber\\
 \begin{array}{ccccc}
     &&\hskip-.7cmj-2&&\\
     &\hskip-.2cm\swarrow&\searrow&\\
     j-3&&&\hskip-.3cm j\\
     &\hskip-.2cm\searrow&\swarrow&\\
     &&\hskip-.7cm j-2&&
     \end{array}&\hbox{$j\equiv2$ (mod 3),}\end{array}\right.
     \ee%
for $j=0$, $1$, \ldots, $L$. Here again, simple subquotient
modules labeled $j'\leq0$ are understood to be zero and can be
omitted. In the reducible modules, the pairs connected by
southwest arrows form standard modules. Given this structure and
the dimensions $2j+1$ for the standard modules, the dimensions of
the simple modules can be found recursively. Introducing
$p=\lfloor j/3 \rfloor$ (where $\lfloor x \rfloor$ is the largest
integer $\leq x$), the dimensions of the simple modules are
$2p+1$, $6p+1$, $4p+4$ for $j\equiv0$, $1$, $2$, (mod 3)
respectively ($p\geq 0$). Thus in terms of dimensions of the
simple subquotients, we have for the two non-trivial classes:
$$
{\rm T}_{3p}:~~~~~\begin{array}{ccccc}
     &&\hskip-.7cm(4p)&&\\
     &\hskip-.2cm\swarrow&\searrow&\\
     (2p-1)&&&\hskip-.3cm(2p+1)\\
     &\hskip-.2cm\searrow&\swarrow&\\
     &&\hskip-.7cm(4p)&&
     \end{array}~~~
{\rm T}_{3p+2}:~~~~~ \begin{array}{ccccc}
     &&\hskip-.7cm(2p+1)&&\\
     &\hskip-.2cm\swarrow&\searrow&\\
     (4p)&&&\hskip-.3cm(4p+4)\\
     &\hskip-.2cm\searrow&\swarrow&\\
     &&\hskip-.7cm(2p+1)&&
     \end{array}~~~~~
     $$
The $q$-dimensions, and the dimensions and superdimensions of the
standard and tilting modules over ${\cal A}_{n+1|n}$, can be
calculated similarly, as we showed in the case of $m=0$. In
particular, the $q$-dimensions (superdimensions) $D_j^+$ of all
the tilting modules over \uqsl2 (resp.,\ ${\cal A}_{n+1|n}$) are
zero except for the $j=0$ tilting module, which has $D_0^+=1$.

\begin{figure}
      \begin{center}
       \leavevmode
       \epsfysize=80mm{\epsffile{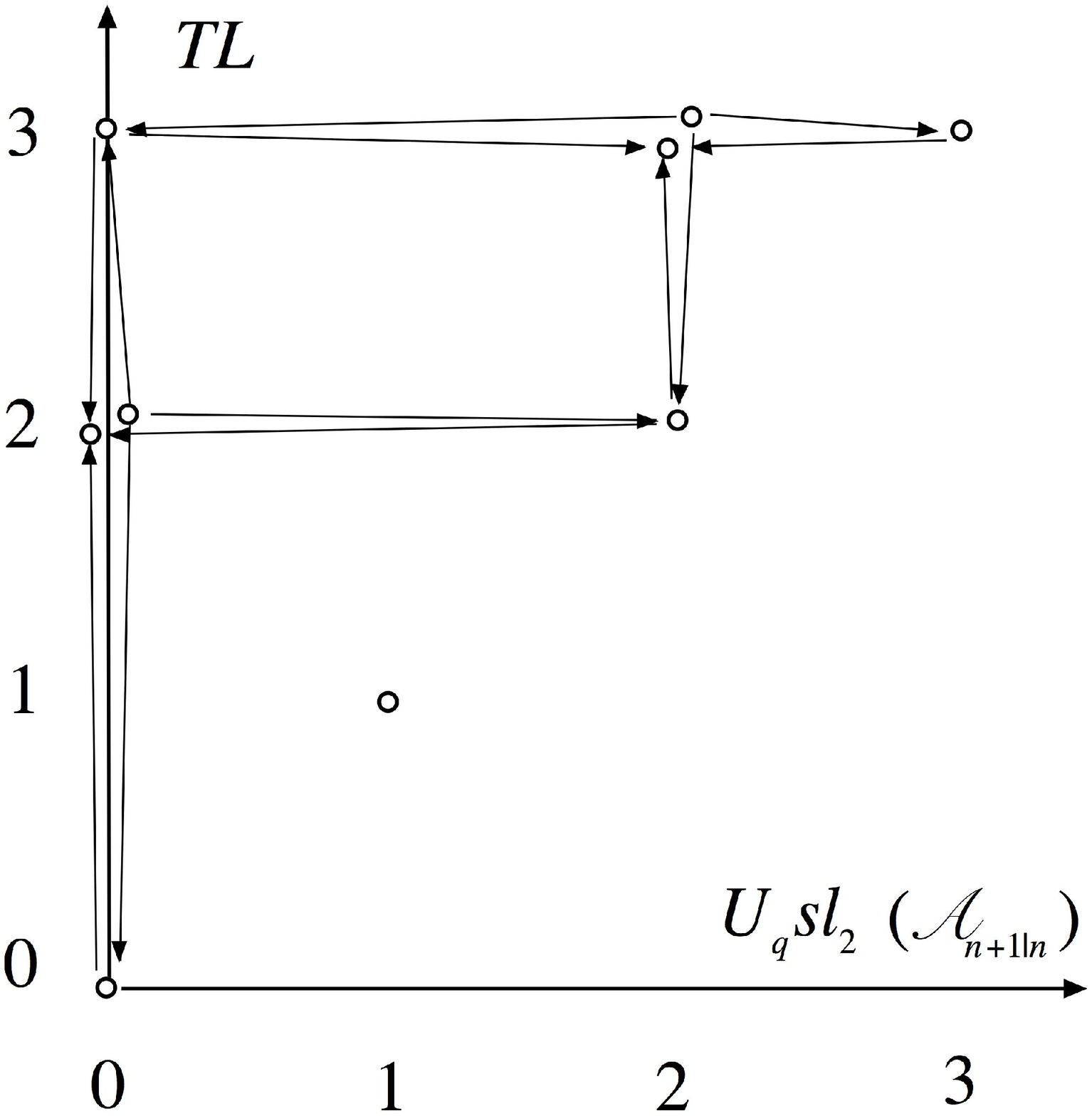}}
       \end{center}
       \protect\caption{The structure of the spin-1/2 chain
       for $q=e^{i\pi/3}$ and $2L=6$ sites, as a representation of
       $U_q($sl$_2)\otimes$TL$_{2L}(q)$, or the same for the
       gl($1+n|n$)-supersymmetric chain with $2L=6$ sites as a representation
       of ${\cal A}_{1+n|n}(2L)\otimes$TL$_{2L}(q)$.}
      \label{nesti}
      \end{figure}

Taking $L=3$ as an example, the decomposition of the Hilbert space
of the chain under \uqsl2 is, in terms of spins $j$,
$$
  0\times 1,~~~1\times 9,~~~
    \begin{array}{ccc}
    0&\\
    &\hskip-.2cm\searrow\\
    &&\hskip-.3cm2\\
    &\hskip-.2cm\swarrow\\
    0&
    \end{array}~~\times 4,~~~~~\begin{array}{ccccc}
    &&\hskip-.7cm2&&\\
    &\hskip-.2cm\swarrow&\searrow&\\
    0&&&\hskip-.3cm3\\
    &\hskip-.2cm\searrow&\swarrow&\\
    &&\hskip-.7cm2&&
    \end{array}~~\times 1,
    $$
for a total of 64 states indeed. The staircase diagram is shown in
Fig.\ \ref{nesti}.

For the TL algebra, the direct summands, with their
multiplicities, are \cite{martinbook,Martini,Martinii}
    \begin{equation}
    \begin{array}{ccccc}
    &&\hskip-.7cm(4)&&\\
    &\hskip-.2cm\swarrow&\searrow&\\
    (1)'&&&\hskip-.3cm(1)\\
    &\hskip-.2cm\searrow&\swarrow&\\
    &&\hskip-.7cm(4)&&
    \end{array}\hskip-.3cm,~(9)~\times 3,\begin{array}{ccc}
      &&\hskip-.3cm(1)\\
      &\hskip-.2cm\swarrow&\\
      (4)&&\\
      &\hskip-.2cm\searrow&\\
      &&\hskip-.3cm(1)
      \end{array}\times 4,~~
   (1)~\times 3,
  \label{TLIindec}
    \end{equation}
in order of increasing $j$. The dimensions of the simple
subquotients are the multiplicities of the \uqsl2 tilting modules,
and the multiplicities are the dimensions of the \uqsl2 simple
modules. But notice that dimension $1$ appears in connection with
two different summands under $U_q($sl$_2)$, and these are not
expected to be isomorphic as TL modules---they have different $j$
values. For this reason, one of them has been distinguished here
by writing $(1)'$. All of these indecomposable modules are
projective, except for the TL singlet at the end. However, there
must also be an additional projective module, which has the simple
module $(1)'$ at the top, and must ``cover'' the lower part of the
TL tilting module in which it occurs. This remaining
indecomposable projective module thus has the form%
\be%
\begin{array}{ccc}
(1)'&&\\
&\hskip-.3cm\searrow&\\
&&\hskip-.3cm(4)\end{array}
\ee%
(and is a standard module), for a total of $L+1=4$ isomorphism
classes as expected here, since they must correspond to the four
distinct indecomposable tilting modules of $U_q($sl$_2)^{(4)}$.

For general $L$, the pattern is similar. There is always a singlet
standard module with spin $j=L$ which is a TL tilting module. For
$L\not\equiv 1$ (mod 3), it is not projective. For $L\equiv 1$
(mod 3), this singlet is projective as well as tilting. But then
there is always a simple summand (tilting module) at the largest
$j\not \equiv 1$ (mod 3), which is not a TL projective module. The
remaining indecomposable tilting modules [including those standard
modules with $j\equiv1$ (mod3) which are simple] are also
projective, and there is one additional projective module for
$j=0$, that has a (different) singlet at the
top. Thus the projective modules for TL are \cite{martinbook,Martini,Martinii},
in terms of spin labels,%
\be%
{\rm P}_{j}:~~~~~\left\{\begin{array}{cl}
\begin{array}{ccc}
0&&\\
&\hskip-.2cm\searrow&\\
&&\hskip-.3cm 2\end{array}&\hbox{$j=0$,}\nonumber\\
\begin{array}{ccccc}
     &&\hskip-.7cm j&&\\
     &\hskip-.2cm\swarrow&\searrow&\\
     j-1&&&\hskip-.3cm j+2\\
     &\hskip-.2cm\searrow&\swarrow&\\
     &&\hskip-.7cmj&&
     \end{array}&\hbox{$j\equiv0$ (mod 3) and $j>0$,}\nonumber\\
     &\nonumber\\
     j&\hbox{$j\equiv1$ (mod 3),}\nonumber\\
     &\nonumber\\
 \begin{array}{ccccc}
     &&\hskip-.7cm j&&\\
     &\hskip-.2cm\swarrow&\searrow&\\
     j-2&&&\hskip-.3cm j+1\\
     &\hskip-.2cm\searrow&\swarrow&\\
     &&\hskip-.7cm j&&
     \end{array}&\hbox{$j\equiv2$ (mod 3),}\end{array}\right.
     \ee%
where again simple modules with $j<0$ or $j>L$ are understood to
be zero. The standard sub- and quotient modules are the pairs
connected by a southeast arrow.

Once this structure is known, the dimensions $d_j^0$ of the simple
modules can be found recursively. For $j\equiv0$, the standard
module is simple, and has dimension $d_j^0=d_j$ as before. For the
other cases, one has $d_j^0=d_j-d_{j+2}^0$ [$j\equiv0$ (mod 3)]
and $d_j^0=d_j-d_{j+1}^0$ [$j\equiv 2$ (mod 3)]. These can be
solved starting from $d_L=d_L^0=1$. It is possible to show that
$d_0^0=1$ for all $L$ (as already stated above), and then the
recursion can also be solved starting from the $j=0$ end. For
$q=e^{i\pi/3}$, we have not found any very illuminating
closed-form expressions for the dimensions $d_j^0$ for $j>0$.

\begin{figure}
  \begin{center}
   \leavevmode
   \epsfysize=80mm{\epsffile{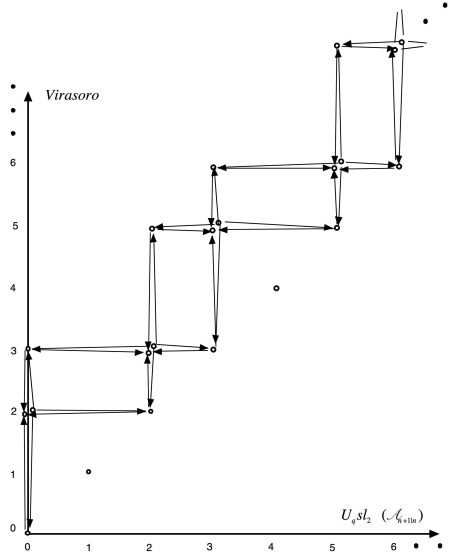}}
   \end{center}
   \protect\caption{Space of states in the continuum $m=1$ theory,
   showing commuting actions of Virasoro and $U_q($sl$_2)$ (or
   ${\cal A}_{1+n|n}$).}
  \label{rep8}
  \end{figure}

In the large $L$ limit, the staircase diagram continues to
infinity, and resembles that for $q=i$, except that extra simple
modules appear at $j\equiv 1$ (mod 3), and there is now a node at
$(0,0)$; see Fig.\ \ref{rep8}.

Let us now consider the continuum limit in more detail. For $c=0$,
the Kac formula reads
\begin{equation}
    h_{rs}=\frac{(3r-2s)^{2}-1}{ 24}.
    \end{equation}
The conformal weights appearing of the primary fields appearing as
subquotients in the lattice partition function are
    \begin{equation}
    h_{1,1+2j}=\frac{j(2j-1)}{3}.
    \end{equation}
We again consider only the  case $j$ integer, and often abbreviate
$h_{1,1+2j}= h_{j}$. For $j\not\equiv1$ (mod 3), these weights are
integers.

Introducing the same notation $K_{1,1+2j}\equiv K_{j}$ for the
Virasoro characters of the standard modules, which we again denote
$r_{1,1+2j}\equiv r_{j}$, we have
    \begin{equation}
    K_{j}\equiv
    K_{1,1+2j}=\frac{\widehat{q}^{h_{1,1+2j}}-\widehat{q}^{h_{1,-1-2j}}}
    {P(\widehat{q})},
    \end{equation}
and from the limit of the spin chain we expect to have the
relations (using again $p=\lfloor j/3\rfloor$)
    \begin{eqnarray}
    K_{3p}&=&\chi_{3p}+\chi_{3p+2},\nonumber\\
    K_{3p+1}&=&\chi_{3p+1},\nonumber\\
    K_{3p+2}&=&\chi_{3p+2}+\chi_{3p+3},\label{identities}
    \end{eqnarray}
where $\chi_j$ are the characters of the associated irreducible
Virasoro representation with highest weight $h_{1,1+2j}$ (denoted
here by $R_{j}$). As for $m=0$, this expectation is correct as it
agrees with the known structure of Virasoro representations in the
Kac table for $c=0$ \cite{BPZ,DFSZ}; the standard modules $r_j$
contain a unique simple submodule when $j\not \equiv 1$ (mod 3).
The irreducible characters are known, and take different forms
depending on $j$ (mod 3) \cite{DFSZ}. If $j=3p+1$, we have
    \begin{equation}
    \chi_{j=3p+1}=\widehat{q}^{(3+12p)^{2}/24}
    \frac{1-\widehat{q}^{3+6p}}
    {P(\widehat{q})}.
    \end{equation}
     If $j=3p+2$, we have
     \begin{equation}
     \chi_{j=3p+2}=
      \sum_{n\notin [-2p-1,-1]}
     \frac{\widehat{q}^{(12n+12p+7)^{2}/24}
     -\widehat{q}^{(12n+12p+11)^{2}/24}}
     {P(\widehat{q})},
     \end{equation}
and if $j=3p$
     \begin{equation}
     \chi_{j=3p}=
      \sum_{n\notin [-2p,-1]}
      \frac {\widehat{q}^{(12n+12p-1)^{2}/24}
      -\widehat{q}^{(12n+12p+7)^{2}/24}}
      {P(\widehat{q})}.
         \end{equation}
In particular, the irreducible character of the identity is
$\chi_{0}=1$.

For $j=3p+1$, the standard representation $r_{j=3p+1}$ is
irreducible, and we can view it as our projective module. For
$j\neq1$ (mod 3) we expect that there are indecomposable
projective modules ${\cal R}_{j}$ that are the limits of the TL
modules, and correspond to them in submodule structure, thus
\be%
{\cal R}_{j}:~~~~~\left\{\begin{array}{cl}
\begin{array}{ccc}
R_0&&\\
&\hskip-.2cm\searrow&\\
&&\hskip-.3cm R_2\end{array}&\hbox{$j=0$,}\nonumber\\
\begin{array}{ccccc}
     &&\hskip-.7cm R_j&&\\
     &\hskip-.2cm\swarrow&\searrow&\\
     R_{j-1}&&&\hskip-.3cm R_{j+2}\\
     &\hskip-.2cm\searrow&\swarrow&\\
     &&\hskip-.7cm{\cal R}_j&&
     \end{array}&\hbox{$j\equiv0$ (mod 3) and $j>0$,}\nonumber\\
     &\nonumber\\
      R_j&\hbox{$j\equiv1$ (mod 3),}\nonumber\\
      &\nonumber\\
 \begin{array}{ccccc}
     &&\hskip-.7cm R_j&&\\
     &\hskip-.2cm\swarrow&\searrow&\\
     R_{j-2}&&&\hskip-.3cm R_{j+1}\\
     &\hskip-.2cm\searrow&\swarrow&\\
     &&\hskip-.7cm R_j&&
     \end{array}&\hbox{$j\equiv2$ (mod 3),}\end{array}\right.
     \ee%
for $j=0$, $1$, \ldots. Apart from the the $j=0$  case, and the
intruding $j\equiv 1$ (mod 3) simple modules, these resemble the
projective modules of the $c=-2$ case. The characters of the
non-trivial (non-standard) cases are clearly, for $j=3p$,
\begin{equation}
    {\rm Tr}_{{\cal
    R}_{j}}\,\widehat{q}^{L_{0}-c/24}
    =K_{3p}+K_{3p-1}=2\chi_{3p}+\chi_{3p-1}+\chi_{3p+2},
    \end{equation}
 and for $j=3p+2$, similarly,
 \begin{equation}
     {\rm Tr}_{{\cal
     R}_{j}}\,\widehat{q}^{L_{0}-c/24}=K_{3p+2}+K_{3p}
     =2\chi_{3p+2}+\chi_{3p}+\chi_{3p+3}.
     \end{equation}
The partition function of the continuum limit of the spin-1/2
chain is again given by%
\be%
Z=\sum_{j=0}^\infty(2j+1)K_j=
\sum_{j=0}^\infty(2j+1)\frac{\widehat{q}^{h_{1,1+2j}}-\widehat{q}^{h_{1,-1-2j}}}
    {P(\widehat{q})}.\ee%
For the $q$-trace instead of the trace, one replaces $2j+1$ by
$[2j+1]_q$, and the resulting formula also gives the
``unmodified'' partition function of the supersymmetric models
with symmetry algebra ${\cal A}_{n+1|n}$ \cite{ReadSal}. For the
partition function of the latter, one instead replaces $2j+1$ by
the dimensions $D_j'=[2j+1]_{q'}$, where $q'+q'^{-1}=2n+1$ for
these $m=1$ models (see Appendix B in ref.\ \cite{ReadSal}).

Understanding the representations ${\cal R}_{j}$ more explicitly
is more difficult than for $m=0$, since we do not have an
explicitly solvable version analogous to the symplectic fermions.
One might think of using the Feigin-Fuchs free field
representation of the theory properly extended by some extra space
to allow for the presence of Jordan cells, along the lines of
\cite{Fjelstad}. We have not done so however, but suspect that the
continuum limits of the present models are different anyhow, since
they involve a single copy of the $h=0$ field. This will be
discussed in more detail when we turn to the case of the torus in
a later paper.

We can now study fusion of the \uqsl2 tilting modules ${\rm T}_j$.
The decomposition of a tensor product of ${\rm T}_{j_1}$ with
${\rm T}_{j_2}$ has to be considered in six cases, as it depends
on $j_1$ and $j_2$ (mod 3). We let $p_1=\lfloor j_1\rfloor/3$,
$p_2=\lfloor j_2\rfloor/3$. For cases in which $j_1$, $j_2
\not\equiv 1$ (mod $3$), we have:
\begin{equation}
    {\rm T}_{3p_1+2}\otimes {\rm T}_{3p_2+2}={\bigoplus
_{r}}'
    \left(2{\rm T}_{3r+2}\oplus\underline{{\rm T}_{3r}}\oplus{\rm
    T}_{3r+3}\oplus{\rm T}_{3r+1}\oplus\underline{{\rm
    T}_{3r+1}}\oplus\underline{{\rm T}_{3r-2}}\oplus
    {\rm T}_{3r+4}\right),
    \end{equation}
where the sum runs over $r=|p_1-p_2|$, $|p_1-p_2|+1$, \ldots,
$p_1+p_2$. Here and in the following, underlined terms are absent
if $r=0$. Similarly
\begin{equation}
    {\rm T}_{3p_1}\otimes {\rm T}_{3p_2}={\bigoplus
_{r}}'
    \left(2{\rm T}_{3r+2}\oplus\underline{{\rm T}_{3r}}\oplus{\rm
    T}_{3r+3}\oplus4{\rm T}_{3r+1}\right),
    \end{equation}
where the sum runs over $r=|p_1-p_2|$, $|p_1-p_2|+1$, \ldots,
$p_1+p_2-1$. Finally, for the mixed fusion
\begin{equation}
    {\rm T}_{3p_1+2}\otimes {\rm T}_{3p_2}={\bigoplus_{r}}'
    \left(2{\rm T}_{3r}\oplus{\rm T}_{3r-1}\oplus{\rm T}_{3r+2}\oplus
    2{\rm T}_{3r-2}\oplus2{\rm T}_{3r+1}\right),
    \end{equation}
where the sum runs over $r=p_1-p_2+1$, $p_1-p_2+2$, \ldots,
$p_1+p_2$ if $p_1\geq p_2$ and $r=p_2-p_1$, $p_2-p_1+1$, \ldots,
$p_1+p_2$ if $p_1< p_2$.

For the cases in which (without loss of generality) $j_1\equiv 1$
(mod 3), we have
\begin{equation}
    {\rm T}_{3p_1+1}\otimes {\rm T}_{3p_2+1}={\bigoplus_{r}}'
    \left({\rm T}_{3r+1}\oplus{\rm
    T}_{3r+2}\right),
    \end{equation}
 where the sum runs over $r=|p_1-p_2|$, $|p_1-p_2|+1$, \ldots, $p_1+p_2$.
 Moreover
 \begin{equation}
     {\rm T}_{3p_1+1}\otimes {\rm T}_{3p_2+2}={\bigoplus
_{r}}'
    \left( 2{\rm T}_{3r+1}\oplus\underline{{\rm T}_{3r}}\oplus{\rm
    T}_{3r+3}\right),
\end{equation}
where the sum  runs over $r=|p_1-p_2|$, $|p_1-p_2|+1$, \ldots,
$p_1+p_2$, and
\begin{equation}
    {\rm T}_{3p_1+1}\otimes {\rm T}_{3p_2}={\bigoplus
_{r}}'
    \left(2{\rm T}_{3r-1}\oplus\underline{{\rm T}_{3r-2}}\oplus{\rm
    T}_{3r+1}\right),
\end{equation}
where the sum runs over $r=p_1-p_2+1$, $p_1-p_2+2$, \ldots,
$p_1+p_2$ if $p_1\geq p_2$ and $r=p_2-p_1$, $p_2-p_1+1$, \ldots,
$p_1+p_2$ if $p_1< p_2$.

Finally, we note the special case $j_1=0$, when
\be%
{\rm T}_0\otimes {\rm T}_j={\rm T}_j\ee%
for all $j\geq 0$.

The fusion rules for the Virasoro projective modules follow
immediately from the correspondence ${\rm T}_j\leftrightarrow{\cal
R}_j$ (and also those for TL projective modules, from the
correspondence ${\rm T}_j\leftrightarrow {\rm P}_j$). As far as we
know, these fusion rules for a logarithmic theory at $c=0$ are
new. This is one of the central results of this paper.

We note that the $j=0$ projective module ${\cal R}_0$ does indeed
behave as the identity in the fusion rules, as expected from the
relation with $U_q($sl$_2)$, because it contains the identity and
the stress tensor, which map any of the projective modules ${\cal
R}_j$ for $j>0$ into themselves. None of the fusion rules produce
the module ${\rm T}_0$ (on the symmetry side) or ${\cal R}_0$ (on
the Virasoro side). The identity field, and the stress tensor,
also occur within the ${\rm T}_2$ (resp., ${\cal R}_2$) modules,
so that conformally-invariant correlation functions do exist.

\section{The case $m=0$ for dilute loops --- critical polymers}

A lattice (vertex) model for dilute loops can be formulated on the
square lattice with a two-colour TL algebra \cite{GrimmPearce}, or
a spin-1 chain with the states $S_z=\pm1$ viewed as occupied by
parts of loops, and $S_z=0$ as empty. It does have \uqsl2
symmetry, with the strands transforming as spin-1/2. From a
symmetry point of view, the models are identical to the dense
cases. This is also true within the supersymmetric formulation
\cite{ReadSal}, in which the obvious supersymmetry algebra is (the
enveloping algebra of) osp($m+2n|2n$), and acts on the occupied
sites as the vector (defining) representation, and trivially on
the empty sites. As the models forbid the strands to cross, the
global symmetry algebra is actually larger than osp($m+2n|2n$)
\cite{ReadSal2}. We will consider here the unoriented version of
dilute loops with $m=0$, for variety and because this is of
natural interest for polymers. This means the number of strands on
any time-slice of the system is unrestricted. Because of the
existence of empty sites, both parities of $2j$ occur for each
length of system. For $m=0$, this model possesses both a
high-temperature phase and a low-temperature phase, separated by a
transition. The continuum limit in the low-temperature phase is in
the same universality class as the (unoriented loops) $m=0$ dense
polymer theory above. The critical point describes dilute
polymers. It is the same universality class as in an
osp($2n|2n$)-invariant $\phi^4$ (Landau-Ginzburg) formulation of
dilute polymers, or as in an osp($2n|2n$)-invariant nonlinear
sigma model with target space a supersphere, $S^{2n-1|2n}$
\cite{ReadSal}.

We will not define in detail here the lattice model, but we make
some comments. For all $q$, the algebra generated by the
nearest-neighbor interactions in the spin-1 chain that represents
dilute loops is clearly somehow similar to the TL algebra, even
though the dimensions of the algebras are different. We suspect
that this relation is in fact Morita equivalence (with a partial
exception that will be mentioned below): the algebra generated by
the spin-1 chain is Morita equivalent to the TL algebra (generated
by the nearest-neighbor terms in the spin-1/2 chain) for the same
number of sites, and at the same $q$. This is motivated by the
common symmetry structure for all $q$, and by the relation of this
with the TL (or Virasoro) structure. It remains a conjecture as we
have not established that the spin-1 model and \uqsl2 form a
``dual pair''. The exception occurs when $m=0$, since as we will
see the singlet module at $(0,0)$ is present in the spin-1 case,
but was absent for the spin-1/2 model. This is connected with a
well-known phenomenon in these loop models: The dilute model
possesses a low-$T$ phase that is in the same universality class
as the dense model. Yet the dilute model has non-vanishing
partition function (here it is the ``unmodified'' partition
function, defined using the $q$-trace in the transfer matrix
approach, that we mean), but that of the dense model vanishes for
$m=0$. The discrepancy is resolved because the state with no
polymer loops exists in the dilute model, and produces the nonzero
partition function, but there is no such configuration in the
dense model. In the dense phase of the dilute model, this state
occurs at high energy and is dropped in the continuum limit, so
the partition function (in which energies of the spin chain are
measured from the ground state energy) becomes zero. This
configuration with no loops corresponds to the state of the chain
at $(0,0)$ in the staircase diagram for $m=0$.

\begin{figure}
  \begin{center}
   \leavevmode
   \epsfysize=80mm{\epsffile{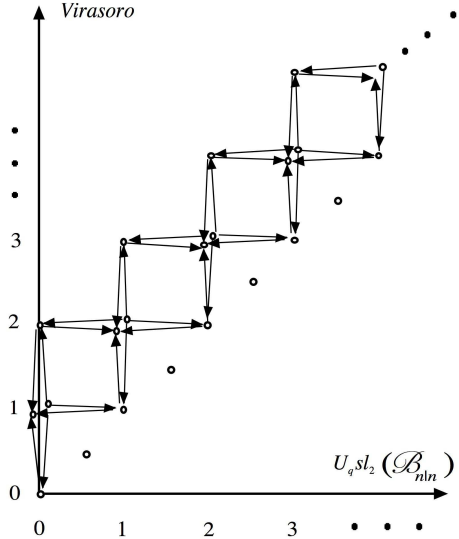}}
   \end{center}
   \protect\caption{Space of states in the continuum dilute $m=0$ theory,
   showing commuting actions of Virasoro and $U_q($sl$_2)$ (or
   ${\cal B}_{n|n}$).}
  \label{polyrep}
  \end{figure}

As we do not consider the lattice algebraic formulation in detail
here, we pass directly to the continuum theory at the critical
point for $m=0$. The staircase diagram is shown in Fig.\
\ref{polyrep}. It has the same form as the dense polymer case,
except that the circle at $(0,0)$ is now included. The \uqsl2
structure is the same as for dense polymers, and we include the
$2j$ odd cases for the unoriented case. Here the singlet tilting
module ${\rm T}_0$ does appear with nonzero multiplicity in the
chain. For the supersymmetric models, the symmetry algebra for the
unoriented loops case is called ${\cal B}_{n|n}$ in Ref.\
\cite{ReadSal2}.

On the Virasoro side, the conformal weights are taken from the
first {\em row} of the $c=0$ Kac table, so
\begin{equation}
    h_{j}\equiv h_{1+2j,1}=\frac{j(3j+1)}{2}.
    \end{equation}
The characters of the standard modules read
\begin{equation}
    K_{j}=\frac{\widehat{q}^{h_{1+2j,1}}-\widehat{q}^{h_{-1-2j,1}}}
    {P(\widehat{q})}.
    \end{equation}
Using standard character formulas we check the identities
\begin{eqnarray}
    K_{p+1/2}&=&\chi_{p+1/2},\nonumber\\
    K_{p}&=&\chi_{p}+\chi_{p+1},
    \end{eqnarray}
similar to the $m=0$ dense  case. The partition function of the
continuum limit of the spin one chain is again given by
\be
  Z=  \sum_{j=0}^\infty(2j+1)\frac{\widehat{q}^{h_{1+2j,1}}-
    \widehat{q}^{h_{-1-2j,1}}}
    {P(\widehat{q})}.\ee
For the $q$-trace instead of the trace, one replaces $2j+1$ by
$D_j=[2j+1]_q$ with $q=i$ (thus $D_j=(-1)^j$ for $j$ integer,
$D_j=0$ for $j$ half-integer), which also gives the unmodified
partition function of the supersymmetric models with symmetry
algebra ${\cal B}_{n|n}$ \cite{ReadSal2}. The latter is simply
$Z_{\rm unmod} =1$ by the Euler identity. For the ``modified''
partition function of the supersymmetric models, one instead
replaces $2j+1$ by the dimensions $D_j'=[2j+1]_{q'}$, where
$q'+q'^{-1}=2n$ for these models (see Appendix B in ref.\
\cite{ReadSal}). For $n=1$, the supersymmetry algebra is
isomorphic to $U_q($sl$_2)$, as in the dense case.

Projective modules for Virasoro are also similar to the dense case
in structure, but now there is also ${\cal R}_0$ for $j=0$, which
is just a standard module, similarly to the $m=1$ (percolation)
case. It has the identity operator at its head, and the stress
tensor $T$ as the highest weight in the foot submodule. These
fields are also part of the ${\cal R}_1$ module, similarly to
percolation ($m=1$).

Fusion rules for $j>0$ are as for dense polymers. In the fusion of
tilting modules of \uqsl2 (or projective modules of Virasoro) for
$j_1$ with $j_2=j_1$ (and $j_1>0$), the term ${\rm T}_0$ (resp.,
${\cal R}_0$) is again absent, even though the corresponding
modules are nonzero; this applies for $2j_1$ odd as well as even.
${\rm T}_0$ behaves as the identity for the fusion rules:%
\be%
{\rm T}_0\otimes {\rm T}_j={\rm T}_j\ee%
for all $j\geq 0$, and similarly for fusion of ${\cal R}_0$ with
${\cal R}_j$. We note that the identity field occurs within the
${\cal R}_1$ module (unlike the dense case), and so
conformally-covariant correlation functions can be produced.


\section{Physical discussion}

In this section, some physical discussion is given, by writing the
algebraic statements in terms of operator products. The discussion
is mainly for $c=0$, and makes contact with that of Gurarie and
Ludwig \cite{GurarieLudwig1,GurarieLudwig2}. We also make some
remarks on the role of fields of weight 1 in the boundary
theories.

The case $c=0$ has probably been the most discussed in the
literature on logarithmic CFTs. There, the existence of the
``$c$-catastrophe'' (i.e., poles in $1/c$ in OPEs as $c\rightarrow
0$) led Gurarie \cite{Gurarie2,GurarieLudwig1,GurarieLudwig2} to
the introduction of a {\sl logarithmic partner} $t$ for the stress
energy tensor, satisfying the OPEs (up to non-divergent  terms):
\begin{eqnarray}
    T(z)T(0)&=&\frac{2T(0)}{z^{2}}+\frac{T'(0)}{z}+\ldots,\nonumber\\
    T(z)t(0)&=&\frac{b}{z^{4}}+\frac{2t(0)+T(0)}{ z^{2}}+\frac{t'(0)}{z}
    +\ldots,\label{tTope}
    \end{eqnarray}
Notice that $T$ is primary, but has a vanishing two-point
function, while $t$ is not primary. The $tt$ OPE itself is more
complicated and starts with a logarithm:
\begin{equation}
    t(z)t(0)=-\frac{2b\ln z}{z^{4}}+\ldots
    \end{equation}
It follows from these equations that $T,t$ form a size two Jordan cell
for $L_{0}$:
\begin{eqnarray}
    L_{0}T&=&2T,\nonumber\\
    L_{0}t&=&2t+T.
   \end{eqnarray}
Gurarie went on to suggest that $T$, $t$ were part of a ``SUSY''
multiplet containing at least two weight-two primary operators
with non-vanishing two-point functions
\cite{Gurarie2,GurarieLudwig1,GurarieLudwig2}.

The foregoing algebraic analysis of our $c=0$ lattice models gives
rise immediately to the existence of $t$. Indeed, for the
continuum limit of the chain, the Virasoro tilting module that
contains the identity and the stress tensor is
\begin{equation}
 {\cal R}_{2}:~~~~~   \begin{array}{ccccc}
      &&\hskip-.7cmR_{2}&&\\
      &\hskip-.2cm\swarrow&\searrow&\\
      R_{0}&&&\hskip-.3cmR_{3}\\
      &\hskip-.2cm\searrow&\swarrow&\\
      &&\hskip-.7cmR_{2}&&
      \end{array}~~~
      \end{equation}
Each simple subquotient module $R_j$ that appears is a highest
weight module, so (by the correspondence of states and operators)
there is a corresponding pseudoprimary field. All these states
have multiplicity 1. Thus the pseudoprimary fields consist of the
identity with conformal weight $h_0=0$, two fields with weight
$h_2=2$, and one with dimension $h_3=5$. One of those with
dimension 2 (the bottom field, as it is primary) is naturally
identified with the stress energy tensor $T$, and should form a
Jordan cell with its logarithmic partner - here the top field
(with conformal weight 2) which we identify with $t$. The whole
indecomposable representation is generated by acting on $t$ with
the Virasoro algebra. We now see that there is a fourth member of
the multiplet of fields, which we call $\tau$, and which has
conformal weight $h_3=5$. We recall that this module can be
analyzed into a submodule that is standard, and which is generated
by the descendants of $1$, and a quotient module that is also
standard, and the quotient module is generated by $t$.

The arrow from $R_{2}$ to $R_{0}$ corresponds to the fact that $t$
is not primary, and can be mapped to the identity under the action
of Virasoro generators - this is directly related with the
existence of the number $b$ in the second OPE (\ref{tTope}). The
arrow from $R_{2}$ to $R_{3}$ indicates that the second OPE in
(\ref{tTope}) must be extended to include the field of dimension
$5$:
\be%
T(z)t(0)\sim
\frac{b}{z^{4}}+\frac{2t(0)+T(0)}{z^{2}}+\frac{t'(0)}{z}+\ldots
+ z c_{Tt}^\tau\tau(0) \ldots\ee%
where $c_{Tt}^\tau$ is some coefficient. $\tau$ is not primary,
but only maps to descendents of $T$ under action of Virasoro with
positive modes:%
\be%
T(z)\tau(0)\sim \frac{c_{T\tau}^T T(0)}{z^5}+\ldots.\ee%
Modulo the Virasoro submodule generated by $T$ in this way, the
field $\tau$ is simply the null Virasoro descendent of $t$ that
occurs in the Virasoro standard module when $c=0$; it must not be
set to zero.

The possible supersymmetry multiplet of fields of weight 2
\cite{Gurarie2,GurarieLudwig1,GurarieLudwig2} may be compared with
the \uqsl2 multiplet structure (horizontal in our diagrams). For
the weight 2 fields for the $m=1$ theory, this is the ${\rm T}_2$
tilting module (or its Morita equivalent in the supersymmetric
theories \cite{ReadSal,ReadSal2}), which has the trio form %
\be%
\begin{array}{ccc}
    0&\\
    &\hskip-.2cm\searrow\\
    &&\hskip-.3cm2\\
    &\hskip-.2cm\swarrow\\
    0&
    \end{array}\ee%
in terms of the spin labels which are $j=0$, $2$ (for the $m=1$
theory). This is in agreement with their expectations, since our
node $j=2$ is a multiplet, not a single field. What plays the role
of supersymmetry in our case is $U_{q}($sl$_{2})$ or the ${\cal
A}_{n+1|n}$ symmetry algebra \cite{ReadSal2}, which is represented
acting horizontally on Fig.\ \ref{rep8}. The dimension of the
space of fields of conformal weight two in our supersymmetric
models turns out to be:
\begin{equation}
   \# \{\hbox{fields with $h=2$}\}=D_{0}'+D_{2}'=(2n+1)^{4}-3(2n+1)^{2}+1
    \end{equation}
(For the limit of the spin-1/2 chain, it is $1+5=6$.) The basic
doublet $t$, $T$ is thus completed by a large number (the
preceding number minus 2, called $D_2'^0$) of other fields with
$h=2$, which themselves lie in the $R_2$ piece of another Virasoro
indecomposable of structure
\begin{equation}
 {\cal R}_{3}:~~~~~   \begin{array}{ccccc}
      &&\hskip-.7cmR_{3}&&\\
      &\hskip-.2cm\swarrow&\searrow&\\
      R_{2}&&&\hskip-.3cmR_{5}\\
      &\hskip-.2cm\searrow&\swarrow&\\
      &&\hskip-.7cmR_{3}&&
      \end{array}.~~~
      \end{equation}
The whole picture is thus considerably more involved than in Ref.\
\cite{Gurarie2,GurarieLudwig1,GurarieLudwig2}.

A similar discussion applies to the dilute polymer case. The
Virasoro tilting module containing the identity and the stress
tensor is then
\begin{equation}
 {\cal R}_{1}:~~~~~   \begin{array}{ccccc}
      &&\hskip-.7cmR_{1}&&\\
      &\hskip-.2cm\swarrow&\searrow&\\
      R_{0}&&&\hskip-.3cmR_{2}\\
      &\hskip-.2cm\searrow&\swarrow&\\
      &&\hskip-.7cmR_{1}&&
      \end{array}~~~
      \end{equation}
and the arrow from $R_{1}$ to $R_{2}$ indicates that in this case
the OPE (\ref{tTope}) must be extended to include a field of
dimension 7, similar to $\tau$; it is a Virasoro descendant that
is a null vector, but must not be set to zero. We have for the
supersymmetric models
\begin{equation}
   \# \{\hbox{fields with $h=2$}\}=D_{0}'+D_{1}'=4n^{2}.
    \end{equation}
For the \uqsl2 theory, this number is $1+3=4$, so that it indeed
turns out to be simply a quartet of fields, and as $q=i$ these can
be viewed as a supersymmetry multiplet also, by putting $n=1$.

It is interesting to observe that reducible but indecomposable
modules of Virasoro of the forms we find are in general
characterized by dimensionless parameters measuring the ``mixing''
of representations, and that the number $b$ of Gurarie
\cite{Gurarie2} appears simply as a particular case of what might
be an infinite set of parameters. Clearly, a more detailed
analysis of indecomposables at $c=0$ along the lines of \cite{Rho}
would be useful here. But it is also important to understand how
features of the lattice models determine the values of these
parameters in our cases.

Gurarie and Ludwig \cite{Gurarie2,GurarieLudwig1,GurarieLudwig2}
also advocated viewing the operator product algebra generated by
$t$ and $T$ as a kind of chiral algebra. But we now see that we
must introduce $\tau$ as a further field, and the fusion rules for
this multiplet with itself produce fields in higher-$j$ projective
modules. That is, they generate by repeated ope's not only the
full set of integer-weight fields, which would then need to be
included in the chiral algebra (and which are not all singlets
under the symmetry algebra), but also for $m=1$ the non-integer
weight fields. In the oriented $m=0$ and $m=1$ theories, this
space of fields corresponds to the full space of states! This
argument ignored the role of the symmetry algebra in the actual
fields, which after all carry representations of the symmetry
algebra also. This might restrict the fields actually produced in
iterated ope's. However, it seems unfair to omit the fields
related to $t$ and $T$ by symmetry from such a chiral algebra. The
fields related by symmetry include some in the ${\cal R}_3$
Virasoro projective module, and fusion of these produces higher
$j$'s. Thus it seems inescapable that all integer $j$ fields,
including those with fractional weights, must be in such a chiral
algebra. In cases such as the dilute unoriented polymers, there
are also fields with $2j$ odd, and all of these together should
probably be viewed as making up a single representation of this
large chiral algebra. (Thus, the structure of the modules over
this larger algebra resembles that of the level-one su(2) affine
Lie algebra theory at $c=1$, of which the \uqsl2 theories are a
deformation.) To us, it does not seem likely to be helpful to try
to view this as a chiral algebra. Instead, the maximal chiral
algebra (in the strict sense of containing only integer-weight
fields) of these boundary theories seems to be the Virasoro
algebra. On the other hand, the symmetry analysis seems powerful.

Finally, we turn to another issue: we point out that neither of
the boundary CFTs with $c=0$ possesses any fields with $h=1$. The
bulk theories in the supersymmetric formulation do possess such
fields of weights $(1,0)$ and $(0,1)$ \cite{ReadSal}, and they
transform in the adjoint representation of the algebras
gl($1+n|n$) and osp($2n|2n$), respectively in the two examples.
The existence of these fields is required by Noether's theorem.
(Despite the conformal weights, these currents are not generally
(anti-)holomorphic, implying that logarithms will appear in their
correlators, and they do not generate affine Lie algebras
\cite{ReadSal}). In the boundary theory, on the other hand, while
there are global symmetries, there is in general no such thing as
a divergenceless boundary current, because charge can escape into
the bulk, and Noether's theorem only implies the existence of bulk
currents. Thus in general there need not be any corresponding
weight-1 fields in the boundary CFT.

On the other hand, a more fruitful point of view about such fields
is that weight one fields on the boundary correspond to marginal
perturbations of the boundary theory, while boundary fields of
weight less than one are relevant perturbations on the boundary.
In particular, exactly marginal perturbations of the boundary
theory correspond to the possibility of a continuous change of
conformally-invariant boundary conditions. For the $c=0$ theories
studied here, we learn that there are no marginal boundary
perturbations, and the only relevant perturbations transform in
representations of the symmetry algebra that do not contain any
singlets. Thus such a perturbation would necessarily break the
global symmetry.

On the other hand, for the $c=-2$ theories, there are a number of
weight 1 fields, transforming in a ${\rm T}_2$ tilting module for
$U_q($sl$_2)$, or a corresponding tilting module for ${\cal
A}_{n|n}$. For symplectic fermions, there are 8 such fields. One
of the fields in the $j=2$ subquotient is the boundary term that
corresponds to the bulk ``topological'' term in the sigma model,
and corresponds to changing the Hall angle in the boundary
condition discussed in Sec.\ 3. This gives a line of boundary
conditions that preserve conformal invariance on the boundary.
However, as we noted above, these theories can be transformed onto
one another by a duality rotation, so this perturbation is
``redundant''. The behavior under all the other perturbations is
currently not known to us.


\section{Conclusion}

The  analysis we have presented can be extended to the case of
other values of $q$ a root of unity, in relation with loop models.
It can also be extended to other boundary conditions for open
chains, which lead generically to standard modules with $h_{r,s}$,
$r>1$ using results from Refs.\ \cite{Blob} and \cite{Nichols};
see also Ref.\ \cite{Zuber}. In all cases, the patterns are quite
similar, in particular the reducible indecomposable projective
Virasoro modules usually have the quartet form.

The real problem now is to extend the construction to the closed
or periodic boundary condition case: we hope to report on this
elsewhere.

While this paper and its companion were being completed, an
interesting preprint appeared on the archive
\cite{PearceRasmussen} which has some overlap with our analysis of
the dense polymers case. In this reference, some of the fusion
rules of the conformal field theory are conjectured based on a
diagrammatic interpretation of fusion within the TL algebra. Our
paper provides the correct algebraic framework to understand the
observations of \cite{PearceRasmussen}, together with the tools
allowing immediate generalization to all roots of unity cases. We
also would like to mention the recent work \cite{Tipunin} where
\uqsl2 at roots of unity is used to build indecomposable
representations of chiral $W$-algebras. As discussed in
\cite{ReadSal} the theories we are considering are not rational
(though most likely quasi-rational) and we do not think the
results of \cite{Tipunin} apply to our case.


\vspace{0.5in}

{\bf Acknowledgments}

N.R. is grateful to the NSF for support under grant number
DMR-02-42949. H.S. thanks V. Schomerus for useful discussions.


\begin{thebibliography}{99}

\bibitem{Lectures} M. Flohr, Int. J. Mod. Phys. A {\bf 18}, 4497 (2003); M.
    Gaberdiel, Int. J. Mod. Phys. A {\bf 18}, 4593 (2003).
\bibitem{Flohr} M. Flohr and A. Mueller-Lohmann, J. Stat. Mech. {\bf 0604},
P002 (2006).
\bibitem{glr} I.A. Gruzberg, A.W.W. Ludwig, and N. Read, Phys.
Rev. Lett. {\bf 82}, 4524 (1999).
\bibitem{Leclair} S. Guruswamy, A. Leclair and A.W.W. Ludwig, Nucl. Phys. B {\bf
583}, 475 (2000).
\bibitem{Caux} M.J. Bhaseen, J.-S.Caux, I.I. Kogan  and A.M. Tsvelik,
Nucl. Phys. B {\bf 618}, 465 (2001).
\bibitem{Volker} G. Gotz, T. Quella and V. Schomerus, ``The WZNW model
on PSU($1,1|2$)'', hep-th/0610070.
\bibitem{Knizhnik} V.G. Knizhnik, Comm. Math. Phys. {\bf 112}, 567 (1987).
\bibitem{RozSal} L. Rozansky and H. Saleur, Nucl. Phys. B {\bf
376}, 461 (1992).
\bibitem{Gurarie1} V. Gurarie, Nucl. Phys. B {\bf 410}, 535 (1993).
\bibitem{Kausch} M.R. Gaberdiel and H.G. Kausch,  Phys. Lett.
B {\bf 386}, 131 (1996); Nucl. Phys. B {\bf 538}, 631 (1999); H.G.
Kausch, Nucl. Phys. B {\bf 583}, 513 (2000).
\bibitem{SchomSal} V. Schomerus and H. Saleur, Nucl. Phys. B {\bf
734}, 221 (2006).
\bibitem{Flohri} M.A.I. Flohr, Int. J. Mod. Phys. A {\bf 11}, 4147 (1996);
{\it ibid.} A {\bf 12}, 1943 (1997).
\bibitem{Feigini} B. L. Feigin, A.M. Gainutdinov, A.M. Semikhatov, and
I.Yu Tipunin, Commun. Math. Phys. {\bf 265}, 47 (2006).
\bibitem{Fjelstad} J. Fjelstad, J. Fuchs, S. Hwang, A.M.
Semikhatov and I. Yu Tipunin, Nucl. Phys. B {\bf 633}, 379 (2002).
\bibitem{Flohrii} M. Flohr and A. M\"uller-Lohmann,
hep-th/0510096.
\bibitem{Gurarie2} V. Gurarie, Nucl. Phys. B {\bf 546}, 765
(1999).
\bibitem{GurarieLudwig1}V. Gurarie and A. W. W. Ludwig, J. Phys.
A {\bf 35}, L377 (2002).
\bibitem{GurarieLudwig2} V. Gurarie and A. W. W. Ludwig,
hep-th/0409105.
\bibitem{Cardy} J. Cardy, ``The stress tensor in quenched random
systems'', in {\it Statistical Field Theories}, Proceedings of the
NATO Advanced Research Workshop on Statistical Field Theories,
June 2001, A. Cappelli and G. Mussardo, Eds, NATO Science series,
Vol.\ II/73 (Kluwer Academic, Dordrecht, 2002).
\bibitem{ReadSal} N. Read and H. Saleur, Nucl. Phys. B {\bf 613}, 409 (2001).
\bibitem{ms} G. Moore and N. Seiberg, Commun. Math. Phys.
{\bf 123}, 177 (1989).
\bibitem{ReadSal2}N. Read and H. Saleur, ``Enlarged symmetry
algebras of spin chains, loop models, and $S$-matrices'',
cond-mat/0701xxx.
\bibitem{Zuber} P. A. Pearce, J. Rasmussen and J.B. Zuber,
``Logarithmic minimal models'', hep-th/0607232.
\bibitem{martinbook} P.P.~Martin, {\it Potts Models and Related Problems
in Statistical Mechanics} (World Scientific, Singapore, 1991).
\bibitem{Martini} P. P. Martin, Int. J. Mod. Phys. A {\bf 7}, Supp.
1B, 645 (1992).
\bibitem{Martinii} P. P. Martin and D. McAnally, Int. J. Mod.
Phys. A {\bf 7}, Supp. 1B, 675 (1992).
\bibitem{Jonesi} V.F.R. Jones, Invent. Math. {\bf 72}, 1 (1983).
\bibitem{KooSal} W.M. Koo and H. Saleur, Nucl. Phys. B {\bf 426}, 459 (1994).
\bibitem{FevePear} G. Feverati and P. Pearce, Nucl. Phys. B {\bf
663}, 409 (2003).
\bibitem{Rho} F. Rohsiepe, ``On Reducible but Indecomposable
Representations of the Virasoro Algebra'', hep-th/9611160.
\bibitem{AndersonFuller} F.W. Anderson and K.R. Fuller, {\it Rings
and Categories of Modules}, 2nd Ed., Graduate Texts in Mathematics
13 (Springer-Verlag, New York, NY, 1992).
\bibitem{cft}For a review, see e.g.\ P. Di Francesco, P. Mathieu,
and D. Senechal, {\it Conformal Field Theory} (Springer, New York,
1997).
\bibitem{BPZ} A.A. Belavin, A.M. Polyakov and A.B. Zamolodchikov,
Nucl. Phys. B {\bf 241}, 333 (1984).
\bibitem{sb} H. Saleur and  M. Bauer, Nucl. Phys. B {\bf 320}, 591
(1989).
\bibitem{cardy}J. Cardy, J. Stat. Phys. {\bf 125}, 1 (2006).
\bibitem{gsw} M.B. Green, J.H. Schwarz, and E. Witten, {\it
Superstring Theory, Volume 2: Loop Amplitudes, Anomalies, and
Phenomenology}, (Cambridge University, Cambridge, 1987), p.\ 204
ff.
\bibitem{nien} B. Nienhuis, Phys. Rev. Lett. {\bf 49}, 1062 (1982).
\bibitem{dfsz} P. di Francesco, H. Saleur, and J.B. Zuber,
J. Stat. Phys. {\bf 49}, 57 (1987).
\bibitem{nienrev} B. Nienhuis, J. Stat. Phys. {\bf 34}, 731
(1984).
\bibitem{PasquierSaleur} V. Pasquier and H. Saleur, Nucl. Phys.
     B {\bf 330}, 523 (1990).
\bibitem{lemma} See Ref.\ \cite{AndersonFuller}, Lemma 29.4, p.\ 324.
\bibitem{xrs} S. Xiong, N. Read, and A.D. Stone, Phys. Rev. B {\bf 56}, 3982
(1997).
\bibitem{Kausch1} H.G. Kausch, ``Curiosities at $c=-2$'',
hep-th/9510149.
\bibitem{Gaberdielindec}  M. Gaberdiel and H. G. Kausch, ``Indecomposable
     fusion products'',  hep-th/9604026.
\bibitem{Gaberdielfusion} M.R. Gaberdiel, Int. J. Mod. Phys. A {\bf 9}, 4619
(1994).
\bibitem{DFSZ} P. di Francesco, H. Saleur and J.B. Zuber, Nucl. Phys.
B {\bf 285}, 454 (1987).
\bibitem{GrimmPearce} U. Grimm and P. Pearce, J.Phys. A {\bf 26}, 7435 (1993).
\bibitem{Blob} P. Martin and H. Saleur, Lett. in Math. Phys.
          {\bf 30}, 189 (1994).
\bibitem{Nichols}A. Nichols, ``The Temperley-Lieb algebra and its
      generalizations in the Potts and XXZ models'', hep-th/0509069.
\bibitem{PearceRasmussen} P. Pearce and J. Rasmussen, ``Solvable
critical dense polymers'', hep-th/0610273.
\bibitem{Tipunin} B. Feigin, A.M. Gainutdinov, A.M. Semikhatov and I.
Yu Tipunin, ``Kazhdan-Lusztig dual quantum group for logarithmic
extensions of Virasoso minimal models'', hep-th/0606506.

\end{thebibliography}
\end{document}